\documentclass[11pt]{article}
\usepackage{hyperref}
\usepackage{amsmath}
\usepackage{mathrsfs}
\usepackage{graphicx}
\usepackage{hyperref}
\usepackage{amssymb}
\usepackage{mathrsfs}
\usepackage{amsthm}
\usepackage{graphicx}
\usepackage{color}

\begin{document}

\title{\large \bf Graceful Exit to Radiation Domination in the Starobinsky Model:\\ A Dynamical Realization}

\author{\normalsize{\sc Bhargabi Saha}\footnote{{\tt bhargabi@iitg.ac.in}}\ \   and {\sc Malay K. Nandy}\footnote{{\tt mknandy@iitg.ac.in} (Corresponding Author)}\\{\normalsize\it Department of Physics, Indian Institute of Technology Guwahati,} {\normalsize\it Guwahati 781 039, India}
}

\date{\normalsize (September 30, 2026)}

\maketitle

\begin{abstract}
The Starobinsky model provides a successful realization of inflation, consistent with current cosmological observations, and serves as a benchmark for assessing alternative inflationary models.   Motivated by its observational success, we investigate within this framework the dynamical evolution of the early Universe.

Incorporating radiation production from scalaron decay, we formulate a coupled system of dynamical equations governing the depletion of the scalaron component and the resulting buildup of radiation, while ensuring exact energy conservation in the expanding background. In the Friedmann equation driving the expansion, both the scalaron and radiation components are allowed to gravitate, thereby accounting for the backreaction of the generated radiation.

Starting with a radiation-free Universe and initial conditions derived from the standard slow-roll approximations, exact numerical evolution of the coupled dynamical system shows that the inflationary dynamics is predominantly dictated by the scalaron sector. Despite an ongoing production of radiation due to scalaron decay, the radiation sector remains subdominant, and the standard slow-roll conditions are preserved throughout inflation, yielding approximately 60 e-folds of inflation.

Following the end of inflation, the scalaron transitions to a damped oscillatory phase, accompanied by a rapid growth of radiation. This dynamical transition provides a {\em graceful exit\/} from inflation and a smooth entry into the reheating regime.  In this regime, the radiation component undergoes a rapid and continuous growth with concurrent depletion of the scalaron energy. As a consequence, the radiation energy density grows significantly and ultimately attains a peak that substantially exceeds the residual scalaron energy density, $\rho_{\rm radiation}\gg\rho_{\rm scalaron}$. This strong inequality indicates the onset of a hot, radiation-dominated phase, in agreement with the {\em Kofman-Yi criterion} for successful reheating.

These results thus demonstrate that the coupled scalaron-radiation dynamics in the expanding background of the Starobinsky model provides a self-consistent realization of {\em graceful exit\/} and a smooth transition to a radiation-dominated cosmological phase consistent with the {\em Kofman-Yi criterion}.\\


\end{abstract}

\pagebreak

\tableofcontents

\section{Introduction}

\subsection{Background}

Quantum effects of matter fields in curved spacetimes can modify the effective dynamics of gravity through a modified Einstein field equation, $R_{\mu \nu}-\frac{1}{2} g_{\mu \nu} R= 8 \pi G \langle T_{\mu \nu} \rangle$ \cite{DAVIES1977,DAVIES-EMT,bunch1977stress,Birrell_Davies_1982}. An important realization is the Starobinsky model \cite{STAROBINSKY1980}, which focuses on higher-curvature contributions arising from the vacuum expectation value of the energy-momentum tensor $T_{\mu\nu}$. In particular, a quadratic Ricci scalar correction $R^2$ in the effective  gravitational action can drive the inflationary epoch of the primordial Universe. This theory can equivalently be expressed as Einstein gravity minimally coupled to an  effective scalar degree of freedom, the {\em scalaron}.

Early inflationary scenarios, beginning with the old-inflation scenario of Guth \cite{guth1981inflationary} and Sato \cite{Sato:1980yn}, relied on a false vacuum decay via a first-order phase transition that caused inhomogeneity through bubble nucleation. The subsequent new-inflation scenarios of Linde \cite{Linde-inflation1982} and Albrecht \& Steinhardt \cite{Albrecht1982A} addressed the inhomogeity problem of old inflation with a second order phase transition but  suffered from fine-tuning issues. Linde's chaotic inflation mechanism \cite{linde1984inflationary,linde1983chaotic} subsequently provided a physically viable scenario in which inflation is driven by the slow-roll evolution of a scalar field, the {\em inflaton}. These developments established the scalar-field paradigm for inflation \cite{dolgov1982baryon, linde-particle, linde1995inflation, linde1988chaotic, linde1985initial}, while the scalaron-driven Starobinsky model provides an important and observationally successful scenario.

The inflationary epoch ends after the slow roll phase and the inflaton field enters into a damped oscillatory phase. Since the Hubble expansion rate decreases sufficiently, decay of the inflaton field into particles is expected to generate a reheating phase of the Universe. Early perturbative models of reheating \cite{dolgov1982baryon,abbott1982particle} were unable to produce sufficient reheating. Only the non-perturbative process of parametric resonance \cite{traschen1990,kofman1994reheating,shtanov1995universe,kofman1996origin,kofman1998reheating} predicted an efficient reheating.

In a non-perturbative framework, Kofman et al. \cite{kofman1997towards} studied the preheating stage after inflation taking the chaotic potential $\frac{1}{2} m^2 \phi^2$ and coupling $g^2 \phi^2 \chi^2$ to analyze particle creation from the decay channel $\phi\rightarrow \chi\chi$. They found that the $\chi$ field follows Mathieu's equation so that $\chi$-bosons are efficiently produced via parametric resonance. Further studies on preheating after inflation have been discussed in \cite{kofman1994reheating, kofman1997towards, Greene-kofman1997, kofman1996nonthermal}.

In the context of the Starobinsky model,
its efficacy was studied by considering a model driven by both inflaton field and quadratic Ricci scalar \cite{Gottlober1991} with the scalar field serving to account for the presence of massive matter fields during inflation \cite{Bruck2015}. They obtained the power spectra at the end of inflation and found remarkable agreement with the Planck 2015 data \cite{Planck2015}, confirming high efficiency of the Starobinsky model. While those calculations were carried out in the Einstein frame, the model was reconsidered in Ref. \cite{Bruck2017}. With this setup, they modeled particle creation as being due to the decay of the scalar field to study the preheating and reheating processes after the end of inflation. They found that decay of the scalar field can be significantly affected as the dynamics is altered by the term quadratic in Ricci scalar which is equivalent to a massive scalaron field. If the scalaron mass is increased with respect to the scalar mass, the reheating process gets prolonged with a higher reheating  temperature, whereas preheating becomes inefficient.

Garcia et al. \cite{Garcia_2023} focused upon various cosmological parameters predicted by two generalizations of the Starobinsky model, namely $R^{2n}$ and $\alpha$-models, which reduce to the original Starobinsky model for $n=1$ and $\alpha=1$. The main focus was to obtain the cosmological observables associated with the slow roll regime with constraints from the reheating epoch via the equation of state parameter. They showed that the generalized Starobinsky models with $n\neq 1$ and $\alpha\neq 1$ predict minor modifications to the results of the original Starobinsky model.

\subsection{Motivation}

While cold inflation requires a distinct reheating phase following the inflationary epoch, warm inflation provides a framework in which inflaton and radiation occur concurrently, thereby eliminating the need for a separate reheating stage. The scenario of warm inflation has been investigated extensively in a variety of inflationary models \cite{Berera1996,Berera2005}. Interestingly, some cold-inflation models that are disfavored by observation can become observationally viable when embedded in the warm-inflationary scenario. More recently, the quartic inflation model and the $\alpha$-attractor model have been investigated in the context of warm inflation \cite{Saha2026,Saha2025}.

On the other hand, within the framework of cold inflation, Linde and collaborators \cite{Kallosh2013a,Kallosh2013b,Linde2014} made significant advances in the development of supergravity generalization of chaotic inflation, as well as superconformal extensions of the Starobinsky and $\alpha$-attractor models. Agreement of these models with cosmological observations has further strengthened the phenomenological appeal of the cold-inflationary framework \cite{Linde2025}.

The Starobinsky model of cold inflation \cite{STAROBINSKY1980} occupies a particularly prominent position owing to its remarkable observational success, making its status as a benchmark to compare and assess other cold inflationary models \cite{Planck2018}. Consequently, apart from its inflationary dyamics, the post-inflationary dynamics of the Starobinsky model is particularly important for investigation. While the inflationary predictions of the model are well established, the transition from the inflationary epoch to the subsequent radiation-dominated era involves a nontrivial reheating mechanism.   In this dynamial mechanism, the scalaron must transfer its stored energy to radiation. Understanding this energy-transfer process is therefore essential for establishing a consistent connection between the successful inflationary phase and the hot radiation-dominated Universe that underlies the subsequent big-bang phase of the Universe.

This energy-transfer process leading to the establishment of radiation domination, however, is by no means trivial. In particular, a successful transition from the inflationary epoch to the radiation-dominated era requires a {\em graceful exit} from inflation. This entails a smooth transition from a scalaron-dominated phase to a radiation-dominated phase that requires a smooth transfer of energy from the scalaron sector to radiation in the {\em dynamical\/} process of reheating. Successful reheating, as Kofman and Yi \cite{kofman_PRD_2005}  emphasized,  requires an almost complete conversion of the inflaton (scalaron) energy into radiation, thereby giving rise to a hot, radiation-dominated universe so that subsequent cosmological processes can take place, such as big bang nucleosynthesis and baryogenesis/leptogenesis. In the present context, this requirement translates into the emergence of a phase where the radiation energy density must become significantly larger than that of scalaron,  expressed as
\begin{equation}
\rho_{\rm radiation} \gg \rho_{\rm scalaron}.
\label{koff}
\end{equation}

A central question, therefore, is whether the Universe can undergo a {\em graceful exit} from the inflationary phase into a hot, radiation-dominated epoch complying with  the {\em Kofman-Yi criterion}~\ref{koff}.

In this work, we investigate this central question within the Starobinsky framework, and show how the required radiation-dominated phase can emerge dynamically from an initially scalaron-only, radiation-free state of the Universe.

To this end, we employ the framework of {\em dynamical reheating}, originally proposed by Albrecht et al.~\cite{albrecht1982reheating}, in which the energy transfer from the Coleman-Weinberg inflaton to radiation is treated dynamically and self-consistently.  We formulate the coupled evolution equations for the scalaron and radiation energy densities, together with the background expansion of the Universe, and solve them numerically. This approach enables us to follow the energy transfer throughout the dynamical evolution and to determine whether radiation eventually comes to dominate over the scalaron, as required by Kofman-Yi criterion \ref{koff} for successful reheating.

More specifically, we treat the entire process as a dynamical system described by a coupled set of nonlinear differential equations governing the evolution of the scalaron and radiation energy densities in the expanding background of the Universe.   Exact energy conservation is maintained so that radiation is produced at the cost of scalaron energy storage; the total energy density in the two sectors satisfies the continuity equation throughout the expanding evolution.   Moreover, the Friedmann equation describing the background expansion is sourced self-consistently by the combined energy density of the scalaron and the created radiation. Thus, rather than being treated as a passive component, the radiation produced contributes to (backreacts on) the background expansion and, in turn, affects the evolution of the scalaron and radiation sectors.

Thus, the essential features of our treatment are: (a) Self-consistent scalaron-radiation-gravity dynamics, (b) Exact energy conservation of the coupled system, and (c) Backreaction of the produced radiation through the Friedmann equation.

Owing to the strongly nonlinear and coupled nature of the dynamical system, a closed-form analytical solution is not feasible. We therefore employ numerical integration to determine the time evolution of the scalaron and radiation energy densities along with the Hubble expansion rate so as to characterize the entire evolution.

Starting with an initially scalaron-only, radiation-free state, our numerical evolution reveals, in the initial phase and throughout the inflationary period, strong scalaron domination depite an ongoing energy transfer from scalaron to the radiation sector, while the progressive radiation growth remains negligible in comparison with the scalaron content.

Once the inflatinary period ends, identified by completion of approximately 60 e-folds of inflation accompanied with the onset of an oscillatory scalaron phase, the radiation component acquires a significantly fast growth rate while the scalaron energy content continues to fall. As a result, the Universe continues to reheat with substantial gain in the radiation component. Ultimately, the radiation density attains a maximum while the residual scalaron component becomes negligibly small in comparison with radiation.

Thus, our scalaron-radiation-gravity-coupled evolution addresses the central question through {\em dynamical realization} of several important aspects:
(i) Graceful exit from inflation after $\sim60$ e-folds,
(ii) Smooth entry into the reheating phase with a high rate of radiation growth,
(iii) Smooth emergence of radiation-domination with a negligible residual scalaron, consistent with the Kofman-Yi criterion \ref{koff}  for successful reheating.
Notably, the Universe is ultimately left with negligible scalaron component, and practically filled with radiation alone so as to connect with the well-established big-bang phase.

\subsection{Layout}

The remainder of the paper is organized as follows. In Section~\ref{starobinsky}, we provide a brief overview of the Starobinsky inflationary model and its formulation in the Einstein frame. Section~\ref{backreact} introduces the modified Friedmann equation incorporating the contribution (backreaction) from radiation production due to scalaron decay. In Section~\ref{bottle}, we examine the implications of the Heisenberg uncertainty principle, which imposes a quantum mechanical constraint on the energy transfer from the scalaron to its decay products (radiation). This section also formulates the rate equations governing the evolution of the radiation and scalaron energy densities, ensuring exact energy conservation. Section~\ref{admissibility} determines the initial conditions for inflation in consistency with the slow-roll approximations satisfying approximately sixty e-folds of expansion during inflation.

In Section~\ref{nld}, we lay out the coupled nonlinear dynamical equations in a form suitable for numerical integration, while Section \ref{sec6} presents the results of numerical integration across both the inflationary and reheating phases. Section \ref{sec5} is devoted to a discussion on the inflation and reheating regimes and their implications in graceful exit and the Kofman-Yi condition \ref{koff} resulting from the Starobinski scenario. Finally, Section~\ref{discussion} concludes the paper with a summary and discussion of the findings.

\section{The Starobinsky Model}
\label{starobinsky}

In the Starobinsky scenario of inflation \cite{STAROBINSKY1980}, a quadratic Ricci scalar term $R^{2}$ in the effective action of the gravitational field successfully drives the primordial cosmic inflation, so that
\begin{equation}
\label{action}
S_{\rm eff}=\frac{1}{2} M_P^2 \int d^4 x\,\sqrt{-g}  \left(R+\frac{R^2}{6 \mu^2}\right),
\end{equation}
where $M_P=(8\pi G)^{-1/2}=2.4 \times 10^{18}$ GeV is the reduced Planck mass and $\mu$ is a mass parameter.

Variation of this effectve action with respect to the metric $g_{\mu \nu}$ leads to the modified field equation, given by
\begin{align}
R_{\mu\nu}-\frac{1}{2}g_{\mu\nu}R+\frac{1}{6\mu^2}\left(-2\nabla_\mu\nabla_\nu R+2RR_{\mu\nu}\frac{}{}\right.
\left.-\frac{1}{2}g_{\mu\nu}R^2+2g_{\mu\nu}\Box R\right)=0.
\end{align}

Crucially, this higher-curvture theory can be mapped via a conformal transformation onto Einstein gravity minimally coupled to an effective scalar field, designated as the {\em scalaron}. Following Whitt \cite{Whitt1984}, we make the conformal transformation,
\begin{equation}
\label{confo}
\tilde g_{\mu \nu}= \left(1+\frac{\Theta}{3\mu^2}\right) g_{\mu \nu} =\Psi g_{\mu \nu},
\end{equation}
so that the conformal Ricci tensor turns out to be 
\begin{align}
\tilde R_{\mu\nu}=R_{\mu\nu}-\frac{1}{3\mu^2 \Psi}\nabla_\mu\nabla_\nu \Theta
+\frac{1}{6\mu^4 \Psi^2}\nabla_\mu\Theta\nabla_\nu \Theta
-\frac{1}{6\mu^2 \Psi}g_{\mu\nu} \Box\Theta.
\end{align}
This leads to the field equation in the conformal spacetime as 
\begin{equation}
\tilde R_{\mu\nu}-\frac{1}{2}\tilde g_{\mu\nu}\tilde R
\displaystyle=\frac{1}{6\mu^4\Psi^2}
\left[\partial_\mu\Theta\partial_\nu\Theta-\frac{1}{2}\tilde g_{\mu\nu}
\left(\tilde g^{\rho\sigma}\partial_\rho\Theta\partial_\sigma\Theta+\mu^2\Theta^2\right)\right].
\end{equation}
Introducing a scalar field $\phi$ defined by
\begin{equation}\label{new}
\Psi = \left(1+\frac{\Theta}{3\mu^2}\right) = e^{\sqrt{\frac{2}{3}}\frac{\phi}{M_P}},
\end{equation}
we can find the equivalent effective action as 
\begin{equation}\label{action2}
\tilde S_{\rm eff}=\int d^4 x\,\sqrt{-\tilde g}  \left[\frac{1}{2}M_P^2\tilde R-\frac{1}{2}\tilde g^{\mu\nu}\partial_{\mu}\phi \partial_{\nu}\phi-V(\phi)\right],
\end{equation}
with the scalar potential 
\begin{equation}
\label{potential}
V(\phi)=\frac{3}{4}\mu^2 {M_P}^2\left(1-e^{-\sqrt{\frac{2}{3}}\frac{\phi}{M_P}}\right)^2.
\end{equation}

Thus the Starobinsky model is equivalent to the Einstein gravity minimally coupled to a scalar field  $\phi$, which is called the {\em scalaron} and the parameter $\mu$ is attributed to its mass.

Thus, the energy density $u(t)$ and pressure $P(t)$ of the scalaron field $\phi(t)$ are given by
\begin{equation}
\label{energy}
\begin{array}{c}
u=\frac{1}{2}\dot{\phi}^2+V(\phi)=\frac{1}{2}\dot{\phi}^2+\frac{3}{4}\mu^2 {M_P}^2(1-e^{-\sqrt{2/3}\phi/{M_P}})^2,\\ \\
P=\frac{1}{2} \dot{\phi}^2- V(\phi)=\frac{1}{2} \dot{\phi}^2-\frac{3}{4}\mu^2 {M_P}^2(1-e^{-\sqrt{2/3}\phi/{M_P}})^2.
\end{array}
\end{equation}

The mass of the scalaron is estimated as 
\begin{equation}
\label{mu}
\mu=1.3 \times 10^{-5} \, M_P
\end{equation}
from CMB anisotropy with respect to spectral index of primordial scalar perturbations and tensor perturbations and observed large-scale structure \cite{Faulkner_2007,Komatsu_2009}.

The scalaron field $\phi$ can couple to any matter field. This can be seen by including a bosonic matter field $\chi$ with the action \ref{action}, so that $S_{\rm tot}= S_{\rm eff} + S_{\rm mat}$, where
\begin{equation}
\label{mat1}
S_{\rm mat} = \int d^4x \, \sqrt{-g} \, \left( -\frac{1}{2} g^{\mu\nu} \, \partial_\mu\chi \, \partial_\nu\chi 
- \frac{1}{2} \, m^2 \chi^2 \right).
\end{equation}

Upon making the conformal transformation \ref{confo}, the bosonic field transforms like $\tilde\chi=\Psi^{-\frac{1}{2}}\chi$, and the transformed matter action takes the form
\begin{equation}
\label{mat2}
\tilde S_{\rm mat}  = \int d^4x \, \sqrt{-\tilde g} \, \left( -\frac{1}{2} \tilde g^{\mu\nu} \, \partial_\mu\tilde\chi \, \partial_\nu\tilde\chi - \frac{1}{2} \, m^2 \, e^{-\sqrt{\frac{2}{3}}\frac{\phi}{M_P}}\, \tilde\chi^2 
 -\frac{1}{\sqrt 6} \frac{\tilde\chi}{M_P} \, \tilde g^{\mu\nu} \, \partial_\mu\tilde\chi \, \partial_\nu\phi
-\frac{1}{12} \, \frac{\tilde\chi^2}{M_P^2} \, \tilde g^{\mu\nu} \, \partial_\mu\phi \, \partial_\nu\phi 
\right).
\end{equation}

Thus, in the Einstein frame (denoted above by an over-tilde), the scalaron field $\phi$ acquires coupling with the bosonic field $\chi$ and it decays mostly via the dominant channel $\phi\rightarrow\chi\chi$, compared to which the other channel $\phi\phi\rightarrow\chi\chi$ is Planck-suppressed.

\section{The Friedmann Equation}
\label{backreact}

In this Section, we  incorporate backreaction of the created radiation in the Friedmann equation.
We consider the Friedmann-Lema\^itre-Robertson-Walker (FLRW) model with the metric

\begin{equation}\label{metric}
ds^2=-dt^2+a^2(t)\left\{\frac{dr^2}{1-kr^2}+r^2(d\theta ^2+\sin^2\theta d\varphi^2)\right\}.
\end{equation}

Due to backrection from the created particles constituting the radiation, the first Friedman equation must incorporate the created radiation energy density $\rho$, in addition to the scalaron energy density $u$, leading to
\begin{equation}
\label{fried1}
H^2 = \frac{1}{3M_P^2} \left(u+\rho\right)
= \frac{1}{3M_P^2} \left[\frac{1}{2}\dot\phi^2+\frac{3}{4}\mu^2 {M_P}^2(1-e^{-\sqrt{2/3}\phi/{M_P}})^2+\rho\right],
\end{equation}
assuming the flat model with the curvature parameter $k=0$.

The second Friedman equation must also be affected by the created radiation energy density $\rho$ with its pressure $p$, giving
\begin{equation}\label{fried2}
\dot H+H^2=-\frac{1}{6M_P^2} \left(u+3P+\rho+3p\right).
\end{equation}
Using the first equality of \ref{fried1}, equation \ref{fried2} reduces to
\begin{equation}\label{fried12}
\dot H=-\frac{1}{2M_P^2} \left(u+P+\rho+p\right).
\end{equation}
In the Starobinsky model \ref{energy}, equation \ref{fried12} becomes
\begin{equation}\label{fried123}
\dot H=-\frac{1}{2M_P^2} \left(\dot\phi^2+\frac{4}{3}\rho\right),
\end{equation}
with creation of ultrarelativistic particles having the equation of state $p=\frac{1}{3}\rho$.

We shall use the standard definition of the slow roll parameter, given by
\begin{equation}\label{slowrolldef}
\epsilon=-\frac{\dot H}{H^2}=\frac{3}{2}\left[\frac{\dot\phi^2+\frac{4}{3}\rho}{\frac{1}{2}\dot\phi^2+\frac{3}{4}\mu^2 {M_P}^2(1-e^{-\sqrt{2/3}\phi/{M_P}})^2+\rho}\right].
\end{equation}

In order to obtain the initial conditions, we shall employ standard slow-roll approximations to \ref{slowrolldef} in the slow roll phase of inflation. We shall use these initial conditions in our numerical work, where we shall use the Friedmann equation \ref{fried1} without imposing any approximations.

\section{Scalaron Decay and Radiation Production}
\label{bottle}
    
The inherent unstable nature of the scalaron through the prominent decay channel $\phi \to \chi\chi$ prompts us to account for its {\em spontaneous} decay even in the inflationary phase. We therefore examine how the radiation field of the Universe grows due to radiation creation from scalaron decay during the slow-roll phase of inflation as well as in the post-inflationary reheating phase.

\subsection{Constrained Production of Radiation}

\label{model-II}

As discussed earlier in Sec.~\ref{starobinsky}, the scalaron $\phi$ can acquire coupling with bosonic fields, leading to the most dominant decay channel $\phi\rightarrow\chi\chi$ (as seen from Eq.~\ref{mat2}). This decay channel imposes a strict constraint on the energy of each resulting $\chi$-particle, {\em restricting\/} it by the scalaron mass $\mu$.

As shown in Ref.~\cite{Starobinsky1984}, the scalaron field leads to an effective expansion law $\overline{a(t)} \sim t^{2/3}$, similar to a universe dominated by nonrelativistic dust. Starobinsky interpreted the corresponding dust particles as ``heavy scalar quasiparticles with rest mass $\mu$ and zero momentum,'' implying that the scalaron behaves as a supermassive stationary particle.

Since the main decay channel is $\phi \to \chi\chi$ (the alternative process $\phi\phi\to\chi\chi$ is suppressed by the Planck scale), a supermassive and stationary scalaron produces two {\em ultra-relativistic} $\chi$ particles (bosons). These two {\em identical} particles must fly in opposite directions, each with equal energy $\varepsilon = \frac{1}{2}\mu$, fixed {\em exactly} as required by the conservation of momentum and energy, respectively.

This means that the value of $\varepsilon$ is fixed by kinematics. Even though the energy $\varepsilon$ is fixed, the {\em direction} of the momentum vector ${\bf p}$ is not fixed. Only its magnitude $|{\bf p}| = \sqrt{p_x^2 + p_y^2 + p_z^2} = \frac{\varepsilon}{c}$ is fixed, while the direction of ${\bf p}$ is distributed isotropically in the isotropic background of the 3-space of FLRW metric.

As the length scale of the Universe was extremely small in the primordial phase, it is reasonable to assume that the Universe was dominated by the rules of quantum mechanics. We therefore employ the Heisenberg uncertainty principle, 
$\Delta x \Delta p_x\geq \frac{1}{2}\hbar$, where the {\em equality\/} sign implies the {\em minimum\/} position uncertainty, ${\Delta x}_{\rm min}= \frac{\hbar}{2\Delta p_x}$.

From the definition $(\Delta p_x)^2=\langle p_x^2 \rangle - \langle p_x \rangle^2$, we can estimate the momentum uncertainty $\Delta p_x$ following a few simple arguments: 
(i) Since the spontaneous decay process ought to be highly random, the momentum distribution should be isotropic. This implies $\langle p_x \rangle=\langle p_y \rangle=\langle p_z \rangle=0$ and $\langle p_x^2 \rangle=\langle p_y^2 \rangle=\langle p_z^2 \rangle$, leading to $\langle \mathbf{p}^2 \rangle=\langle p_x^2 + p_y^2 + p_z^2 \rangle = 3\langle p_x^2 \rangle = 3(\Delta p_x)^2$. 
(ii) Since the $\chi$-particles produced from the massive scalaron decay are ultrarelativistic, we can approximate them as massless, giving the dispersion relation $\varepsilon^2=\langle \mathbf{p}^2 \rangle c^2$.
(iii) Thus, using the preceding two results, the momentum uncertainty can be expressed as $\Delta p_x=\frac{\varepsilon}{\sqrt{3}c}$. 

We thus have ${\Delta x}_{\rm min}= \frac{\hbar}{2\Delta p_x}= \frac{\sqrt{3}\hbar c}{2\varepsilon}$, providing the minimum position uncertainty ${\Delta x}_{\rm min}$ in terms of $\varepsilon$.
This position uncertainty makes the configuration space grainy so that we may imagine dividing it into cells each of volume $({\Delta x}_{\rm min})^3$.

Thus, a bound on the radiation energy density naturally arises from the above considerations.
Since the volume $(\Delta x_{\rm min})^3$ appears in the denominator, the {\em maximum} creation of radiation energy density in the decay process is given by
\begin{equation}
w_{\rm max}=\frac{\varepsilon}{({\Delta x}_{\rm min})^3}= \frac{8 \varepsilon}{3\sqrt{3}} \left(\frac{\varepsilon}{\hbar c}\right)^3.
\end{equation}

Correspondingly, maximum possible production rate of radiation energy density is given by the product of the scalaron decay rate $\Gamma$ and the maximal radiation energy density:
\begin{equation}
\label{lambda}
q_{\rm max}=\Gamma \cdot w_{\rm max}=\Gamma \frac{8 \varepsilon}{3\sqrt{3}}\left(\frac{\varepsilon}{\hbar c}\right)^3= \frac{8}{3\sqrt{3}}\Gamma \varepsilon ^4,
\end{equation}   
with $\hbar=c=1$.

Consequently, even if the scalaron field contains a large amount of energy density, the uncertainty principle, in conjunction with momentum and energy conservation in the decay process $\phi\to\chi\chi$, constrain how much of that energy density can be converted into radiation energy density.

In addition, energy conservation demands that the rate of radiation creation \ref{lambda} must equal the rate of loss in scalaron energy density.  This effectively imposes the same limit $q_{\rm max}$ on the energy transfer rate from the scalaron field to radiation.

In the beginning phase ($t\sim t_i$), the scalaron field $\phi$ carries an enormous amount of energy density $(u\gg w_{\rm max})$ so that it can decay into enough number of $\chi$-particles to populate all cells in the entire Universe. As time progresses, the scalaron keeps decaying due both to continual creation of $\chi$-particles from scalaron decay and the ongoing Hubble expansion. Consequently, after sufficient decay, $u$ will decrease to the value $w_{\rm max}$. We designate this time as $t_*$, the instant when $u=w_{\rm max}$. From this time onwards (i.e., for $t>t_*$), the energy density $u$ remaining in the scalaron field would be insufficient $(u< w_{\rm max}$) so that the produced $\chi$-particles would be insufficient to populate all cells in the Universe. Thus, from the time $t_*$ onward, only a fraction $\xi(t)=\frac{u}{w_{\rm max}}$ of the cells can be populated. The production rate of radiation energy density therefore becomes
\begin{equation}
\label{lambda1}
q(t)=q_{\rm max}\xi(t) = \frac{8}{3\sqrt{3}}\Gamma \varepsilon ^4\xi(t),
\end{equation}   
with
\begin{equation}
\label{xi}
\xi(t)=\left\{
\begin{array}{ll}
 1, & \mbox{for } 0\leq t\leq t_*,\\
 \frac{3\sqrt{3}}{8}\frac{u}{\varepsilon^4}, & \mbox{for } t>t_*.
\end{array} \right.
\end{equation}

\color{black}

\subsection{Dynamical Equations}

The constrained production rate $q(t)$ given by \ref{lambda1}, with the bound $q_{\rm max}$ given by \ref{lambda}, acts as a source for the build-up of radiation energy density $\rho(t)$, whose rate of change is governed by the rate equation
\begin{equation}
\label{density1}
\dot{\rho}+3H(\rho+p)=q(t).
\end{equation}
Importantly, the radiation energy density $\rho(t)$ is expected to grow due to the source term $q(t)$, in spite of the prevalent Hubble expansion. 

Since this gain in radiation occurs at the cost of the scalaron, the rate equation for the scalaron energy density $u(t)$ must have a sink term $-q(t)$, in addition to the ambient depletion due to Hubble expansion, so that 
\begin{equation}
\label{udot}
\dot{u}+3H(u+P)=-q(t).
\end{equation}

Thus, the quantity $q(t)$ serves as the transfer rate of energy density from the scalaron field to radiation. This transfer rate is naturally constrained since $q(t)$ is bounded by $q_{\rm max}$, as given by \ref{lambda}.

In conjunction with the Friedmann equation \ref{fried1},
equations \ref{density1} and \ref{udot} define the dynamics of $u(t)$ and $\rho(t)$ in both inflationary and post-inflationary reheating regimes. Together, they imply exact conservation of the total energy, since they give the equation of continuity,
\begin{equation}
\dot\rho_{\rm tot}+3H(\rho_{\rm tot}+p_{\rm tot})=0,
\end{equation}
 where $\rho_{\rm tot}= u+\rho$ and $p_{\rm tot}=P+p$. This confirms that the radiation energy density $\rho(t)$ grows at the cost of decay of the scalaron energy density $u(t)$, respecting strict energy conservation.

Using \ref{energy}, the dynamics of the scalaron field $\phi(t)$ is obtained from Equation \ref{udot} as $\dot{\phi}\ddot{\phi}+3H\dot{\phi}^2+V'(\phi)\dot{\phi}= -q(t)$, or
\begin{align}
\label{phi2}
\dot{\phi}\ddot{\phi}+3H\dot{\phi}^2+\sqrt{\frac{3}{2}}\mu^2 M_P\left\{\left(1-e^{-\sqrt{2/3}\phi/{M_P}}\right)\right.
\left. e^{-\sqrt{2/3}\phi/{M_P}}\right\}\dot{\phi}= -q(t).
\end{align}

In addition, the Hubble expansion rate of the Universe is governed by {\em both} the scalaron and radiation energy densities $u$ and $\rho$ through the Friedmann equation \ref{fried1} in order to account for the backreaction from the created radiation.

\section{Initial Conditions}
\label{admissibility}

The scalaron energy density in the beginning of inflation is extremely high. Consequently, the production rate of energy density of $\chi$-particles is given by $q=\frac{8}{3\sqrt{3}}\Gamma \varepsilon ^4$, as seen from equations \ref{lambda1} and \ref{xi}. Thus the source and sink terms in equations \ref{density1} and \ref{udot} are constant in this phase.

Employing the slow roll condition $\ddot{\phi}\ll H\dot{\phi}$ to Eq. \ref{phi2}, we get 
\begin{equation}\label{phi3}
3H\dot{\phi}^2+\sqrt{\frac{3}{2}}\mu^2 M_P\left\{\left(1-e^{-\sqrt{2/3}\phi/{M_P}}\right)e^{-\sqrt{2/3}\phi/{M_P}}\right\}\dot{\phi}+q=0.
\end{equation}

Moreover, $u\gg \rho$ during inflation, and applying the slow roll approximation $\dot{\phi}^2\ll V(\phi)$ to \ref{fried1} gives $H\approx
\frac{1}{2}\mu \left(1-e^{-\sqrt{2/3}\phi/{M_P}}\right)$. Consequently, equation \ref{phi3} approximates to
\begin{equation}\label{}
\frac{3}{2}\mu \left(1-e^{-\sqrt{2/3}\phi/{M_P}}\right)\dot{\phi}^2
+\sqrt{\frac{3}{2}}\mu^2 M_P\left\{\left(1-e^{-\sqrt{2/3}\phi/{M_P}}\right)e^{-\sqrt{2/3}\phi/{M_P}}\right\}\dot{\phi}+q=0.
\end{equation}
This is a quadratic equation in $\dot\phi$, immediately giving
\begin{equation}
\label{phi41}
\frac{\dot{\phi}}{\frac{\sqrt{3}}{2}\mu M_P}=
\frac{\sqrt{2}}{3}e^{-\sqrt{2/3}\phi/{M_P}}\left[-1\pm \sqrt{1-4\frac{q}{\mu^3 M_P^2}\frac{e^{2\sqrt{2/3}\phi/{M_P}}}{(1-e^{-\sqrt{2/3}\phi/{M_P}})}}\right].
\end{equation}

In addition, the approximation $\rho\ll\dot\phi^2$ should hold in the inflationary regime. We have verified this approximation from our numerical data that indicate the ratio $\frac{4\rho}{3\dot\phi^2}$ grows from zero to $6.8504\times 10^{-11}$ for $\Gamma=\mu$ (and to $7.7303\times 10^{-16}$ for $\Gamma=10^{-5}\mu$) at the end of inflation.

Thus, the slow roll parameter given by Eq. \ref{slowrolldef} can be approximated as $\epsilon=\frac{3\dot\phi^2}{2V}$ in the slow roll approximation $\dot{\phi}^2\ll V(\phi)$, with $V$ given by \ref{potential}. Identifying the end of inflation with $\epsilon=1$ and substituting $\dot \phi_f^2=\frac{2}{3} V_f$ in Eq. \ref{phi41} gives, for the upper and lower ($\pm$) signs, 
\begin{align}
{x_f}^2 (1-x_f)^4=\frac{4}{3}\delta^2\label{cubic21}\\
\frac{4}{3}{x_f}^2 \left[1-\frac{2\delta}{{x_f}^2 (1-x_f)}\right]=(1-x_f)^2,\label{cubic2}
\end{align}
where $x_f=e^{-\sqrt{\frac{2}{3}}\frac{\phi_f}{M_P}}$ and $\tilde{\phi}_f=\frac{\phi_f}{M_P}$ and $\delta=\frac{q}{\mu^3M_P^2}$ are dimensionless quantities. Here we have used the fact that $\delta$ is a very small quantity which can be estimated from 
\begin{equation}
\label{qbar}
\delta=\frac{8}{3\sqrt{3}}\frac{\Gamma\varepsilon^4}{\mu^3 M_P^2}=\frac{\Gamma}{\sqrt{\lambda} M_P}\frac{\varepsilon^4}{\lambda M_P^4},
\end{equation}
where $\lambda=\frac{3}{4}\frac{\mu^2}{M_P^2}$ is introduced in the second equality for later use. From the first equality, we get $\delta=1.6262 \times 10^{-11}$ for $\Gamma=\mu$ (and it takes a still smaller value of $\delta=1.6262 \times 10^{-16}$ for $\Gamma=10^{-5}\mu$). In obtaining these estimates, we have taken the most dominant channel of scalaron decay ($\phi\rightarrow\chi\chi$) so that energy per $\chi$-particle is $\varepsilon= \frac{\mu}{2}$ as given by momentum and energy conservation, and the scalaron mass $\mu$ is taken from Eq. \ref{mu}.

Using the fact that $\delta$ is very small, we approximately find the three roots of equation \ref{cubic2}, among which the only physical (non-zero real-valued) root is 
\begin{equation}
\label{phi--f}
x_f=\left(-3+2\sqrt{3}+\frac{\delta}{-3+2\sqrt{3}}\right),
\end{equation}
whereas we have neglected equation \ref{cubic21} as it gives unphysical solutions.

The number of e-folds $N$ can be obtained from
\begin{equation}
\label{eefold}
N=\int_i^f H dt\approx\frac{1}{2}\mu\int_i^f\left(1-e^{-\sqrt{2/3}\phi/{M_P}}\right)\frac{d\phi}{\dot{\phi}},
\end{equation}
with approximations $\rho\ll\dot\phi^2$ and $\dot\phi^2\ll V(\phi)$ in the slow roll phase.

Expanding Eq.~\ref{phi41} in the leading order in $\delta$, we get
$\tilde{\phi}_f'=-\frac{2\sqrt{2}}{3}\frac{\delta}{x_f(1-x_f)}+O(\delta^2)$
and $\tilde{\phi}_f'=\frac{2\sqrt{2}}{3}x_f\left[-1+\frac{\delta}{x_f^2(1-x_f)}\right]+O(\delta^2)$
for the upper and lower ($\pm$) signs.
Here $\tilde \phi'= \frac{d\tilde \phi}{d\tau}$ and $\tau=\sqrt{\lambda}M_Pt$.
 However, the first expression yields an unphysical value. Substituting the second expression in equation ~\ref{eefold} yields  
\begin{equation}
\label{phi-i1}
\left[\frac{1}{x_i}+\ln x_i+\frac{\delta}{3 x_i^3}\right]=\left[\frac{1}{x_f}+\ln x_f+\frac{\delta}{3 x_f^3}\right]+\frac{4N}{3}.
\end{equation} 
Using Eq.~\ref{phi--f}, we get from Eq.~\ref{phi-i1}
\begin{equation}
\left[\frac{1}{x_i}+\ln x_i+\frac{\delta}{3 x_i^3}\right]=81.3871 -2.0264\,\delta
\end{equation} 
for $N=60$. This gives the initial value, $x_i=e^{-\sqrt{2/3}\phi_i/M_P}$, as
\begin{equation}\label{x_i}
x_i=0.01165+28.9507\,\delta
\end{equation}

Since $\delta$ is very small ($\sim 10^{-11}$ or smaller), we choose $x_i\approx 0.01165$ from \ref{x_i}, giving the initial condition $\tilde\phi_i\approx 5.4531$.  The initial condition on the slope is obtained from Equation \ref{phi41}, yielding
$\tilde\phi_i'=-0.0110$ in the leading order. Thus, using \ref{potential}, the initial condition on the scalaron energy density is found to be $\tilde u_i=\frac{u_i}{\lambda M_P^4}= 0.9769$, in the leading order. It is important to note that these initial conditions follow from the slow roll conditions for inflation with $N=60$ e-folds.

Starting from a radiation-free Universe, we numerically integrate the coupled set of nonlinear differential equations, Eqs. \ref{density1}, \ref{udot}, and \ref{fried1}, using the initial conditions outlined above, supplemented with the source/sink provided by Eqs. \ref{lambda1} and \ref{xi}. This allows us to initiate and sustain the inflationary phase with radiation production, while remaining consistent with the standard slow-roll conditions. Since radiation production is already implemented in these equations, they allow us to directly pass over to the post-inflationary reheating phase.

\section{Coupled Nonlinear dynamics}
\label{nld}

We thus have to integrate three dynamical equations: (i) Rate of depletion in energy density $u$ stored in the scalaron field, namely, Eq.~\ref{udot}, (ii) Rate of gain in radiation energy density $\rho$, Eq.~\ref{density1}, and (iii) the Friedmann equation governed by $u$ and $\rho$ both,  Eq.~\ref{fried1}, with $\rho$ accounting for the backreaction of the created radiation on the expansion dynamics of the Universe.

Upon using the expressions for the scalaron energy density $u$ and pressure $P$ from Eq.~\ref{energy}, these equations can be written as a set of
three coupled differential equations in three variables, $\{u(t),\rho(t),\phi(t)\}$, given by 
\begin{equation}
\label{u-til}
\dot{u}=-3\sqrt{\frac{u+\rho}{3M_P^2}}\,\,\left[2u-\frac{3}{2}\mu^2{M_P}^2(1-e^{-\sqrt{2/3}\phi/{M_P}})^2\right]-q(t),
\end{equation}
\begin{equation}
\label{rho-til}
\dot{\rho}=-4\sqrt{\frac{u+\rho}{3M_P^2}}\,\,\rho+q(t),
\end{equation}
and
\begin{equation}
\label{phi-til}
\dot{\phi}^2=2u-\frac{3}{2}\mu^2{M_P}^2(1-e^{-\sqrt{2/3}\phi/{M_P}})^2,
\end{equation}
where the sink and source terms $\mp q(t)$ are given by Eqs.~\ref{lambda1} and \ref{xi}, that account for the effect of scalaron decay creating radiation with the equation of state $p=\frac{1}{3}\rho$.

The strongly nonlinear and coupled nature of the differential equations, Eqs. \ref{u-til}, \ref{rho-til}, and \ref{phi-til}, precludes the possibility of obtaining an analytical solution. We therefore employ numerical integration to obtain precise, quantitative results. The integration is initialized using values derived from the slow-roll conditions detailed in Section \ref{admissibility}, assuming an {\em initially radiation-free} Universe. Radiation is generated during evolution due to the source term $q(t)$ in Eq.~\ref{rho-til}. We continue the numerical integration beyond the end of inflation so as to smoothly pass  over  into the post-inflationary reheating phase.

In performing the numerical integration, we use the following dimensionless quantities:
\begin{equation}
\label{}
\tau=\sqrt{\lambda}M_Pt, \qquad \tilde{\phi}=\frac{\phi}{M_P}, \qquad \tilde{u}=\frac{u}{\lambda M_P^4}, \qquad \tilde{\rho}=\frac{\rho}{\lambda M_P^4},
\end{equation}
so that the above set of coupled differential equations, Eqs. \ref{u-til}, \ref{rho-til}, and \ref{phi-til}, are expressed in the following dimensionless forms:
\begin{equation}
\label{u-tilde}
\frac{d\tilde{u}}{d\tau}=-3\sqrt{\frac{\tilde{u}+\tilde{\rho}}{3}}\,\,\left[2\{\tilde u-(1-e^{-\sqrt{2/3}\tilde \phi})^2\}\right]-\tilde{q}(\tau),
\end{equation}
\begin{equation}
\label{rho-tilde}
\frac{d\tilde{\rho}}{d\tau}=-4\sqrt{\frac{\tilde{u}+\tilde{\rho}}{3}}\,\,\tilde{\rho}+\tilde{q}(\tau),
\end{equation}
and
\begin{equation}
\label{phi-tilde}
\left(\frac{d\tilde{\phi}}{d\tau}\right)^2=2[\tilde u-(1-e^{-\sqrt{2/3}\tilde \phi})^2],
\end{equation}
where the source/sink terms are dimensionalized as $\tilde q(\tau)=\frac{q(t)}{\lambda^{3/2}M_P^5}$.

In order to confirm the general trend of the dynamical evolution, we carry out the numerical integration of the coupled differential equations, Eqs.\
\ref{u-tilde}, \ref{rho-tilde}, and \ref{phi-tilde},
with six values of $\Gamma/\mu$, namely, $\Gamma/\mu=1,10^{-1},10^{-2},10^{-3},10^{-4},10^{-5}$,
giving six different sets of results.

In discussing our  numerical integration in the next Section, we shall note the results only for $\Gamma=\mu$ and $\Gamma=10^{-5}\mu$, while we present the results for all six values of $\Gamma/\mu$ in tabular forms. For these values of $\Gamma/\mu$, we note that $\tilde q(0)= \left(\frac{4}{3}\right)^{3/2}\delta$ takes the values $\tilde q(0)=2.5037\times 10^{-11}$ and $\tilde q(0)=2.5037\times 10^{-16}$, respectively.

\section{Results of Numerical integration}
\label{sec6}

Corresponding to each of the six values of $\Gamma/\mu$, we numerically integrate the set of coupled nonlinear differential equations \ref{u-tilde}, \ref{rho-tilde}, and \ref{phi-tilde}. To begin the inflationary epoch, we apply the initial conditions $\tilde{\phi}_i=5.4531$, $\tilde{u}_i=0.9769$, $\tilde{\rho}_i=0$ at $\tau=0$, specifically obtained from the standard slow roll conditions with $60$ e-folds, in Section \ref{admissibility}.

  \begin{figure*}[t]
\includegraphics[scale=0.35]{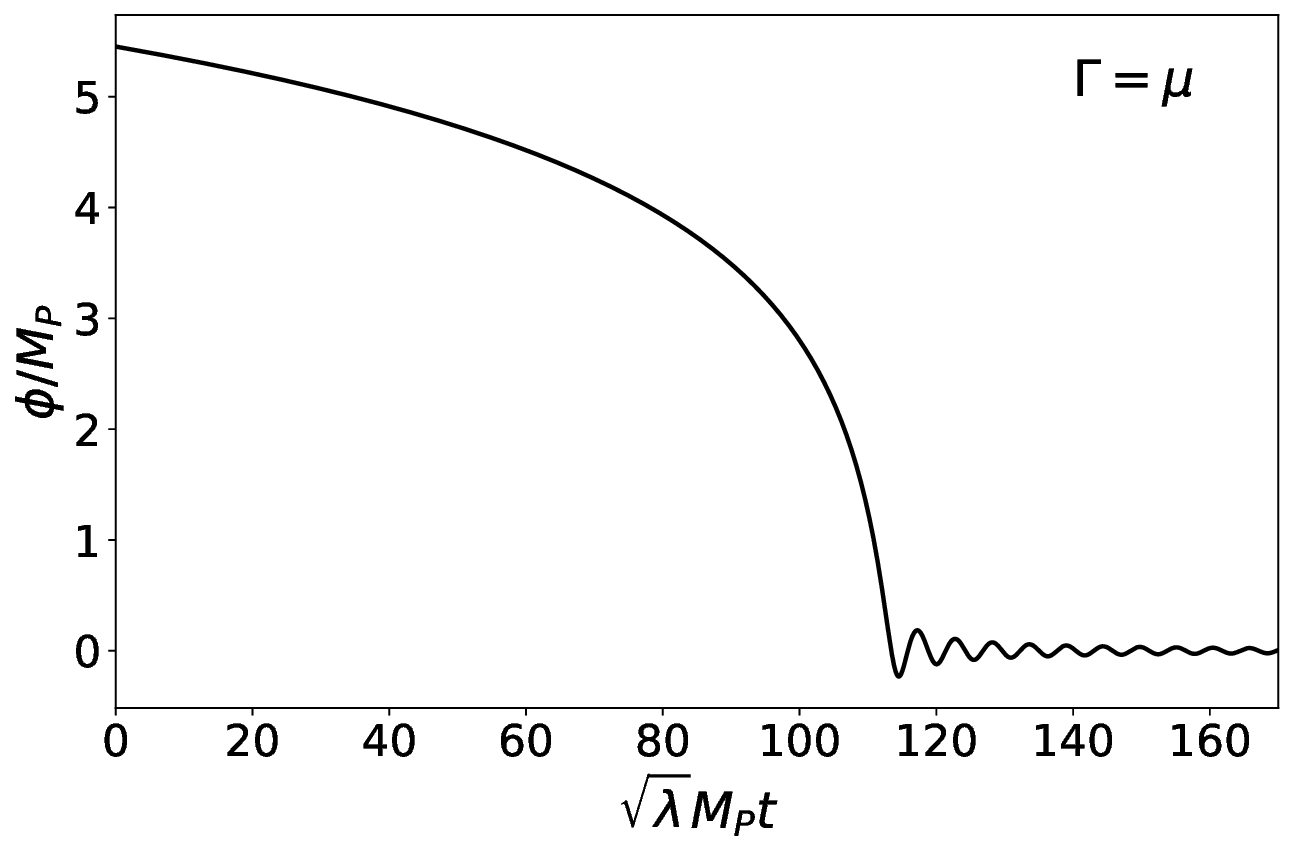}
\includegraphics[scale=0.35]{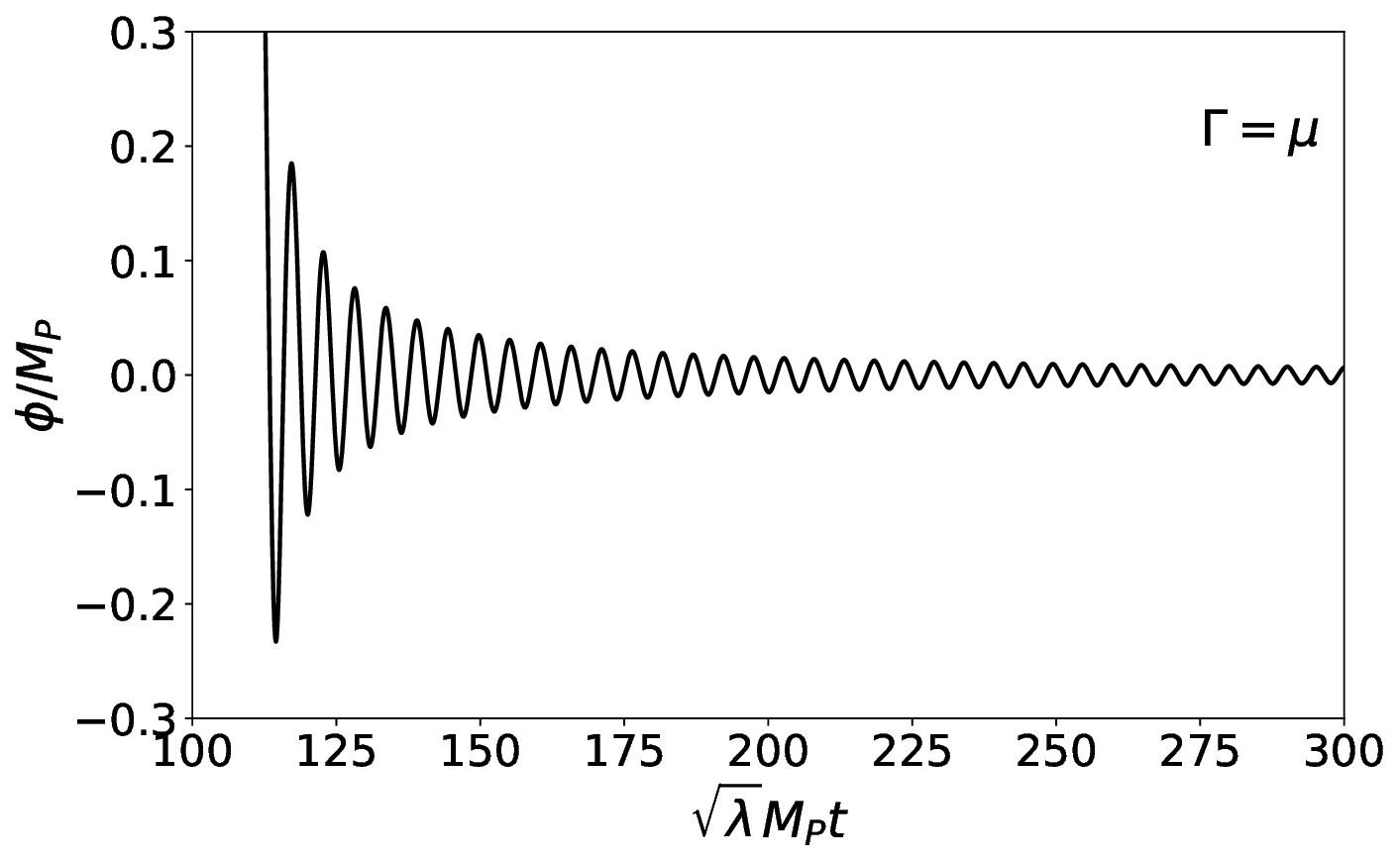}
\includegraphics[scale=0.35]{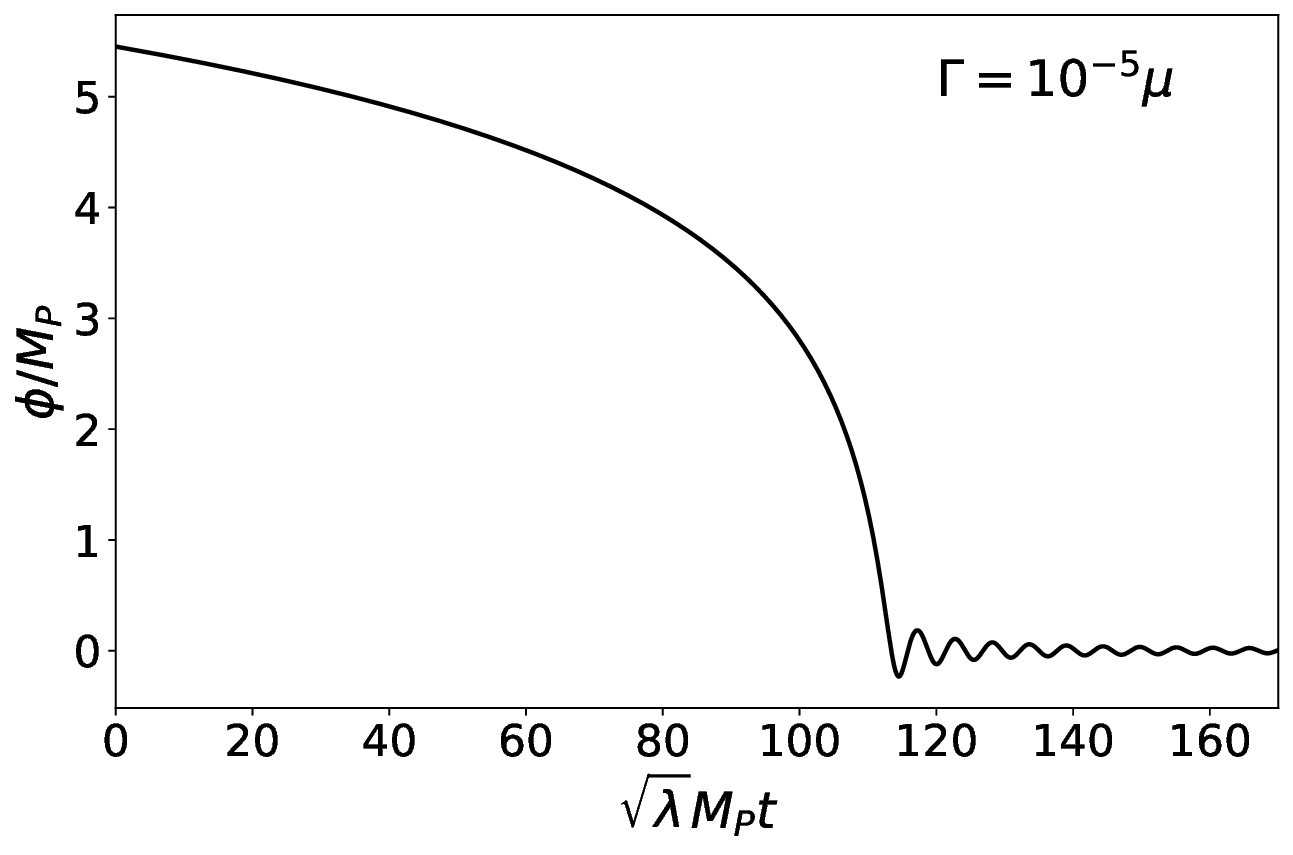}
\includegraphics[scale=0.35]{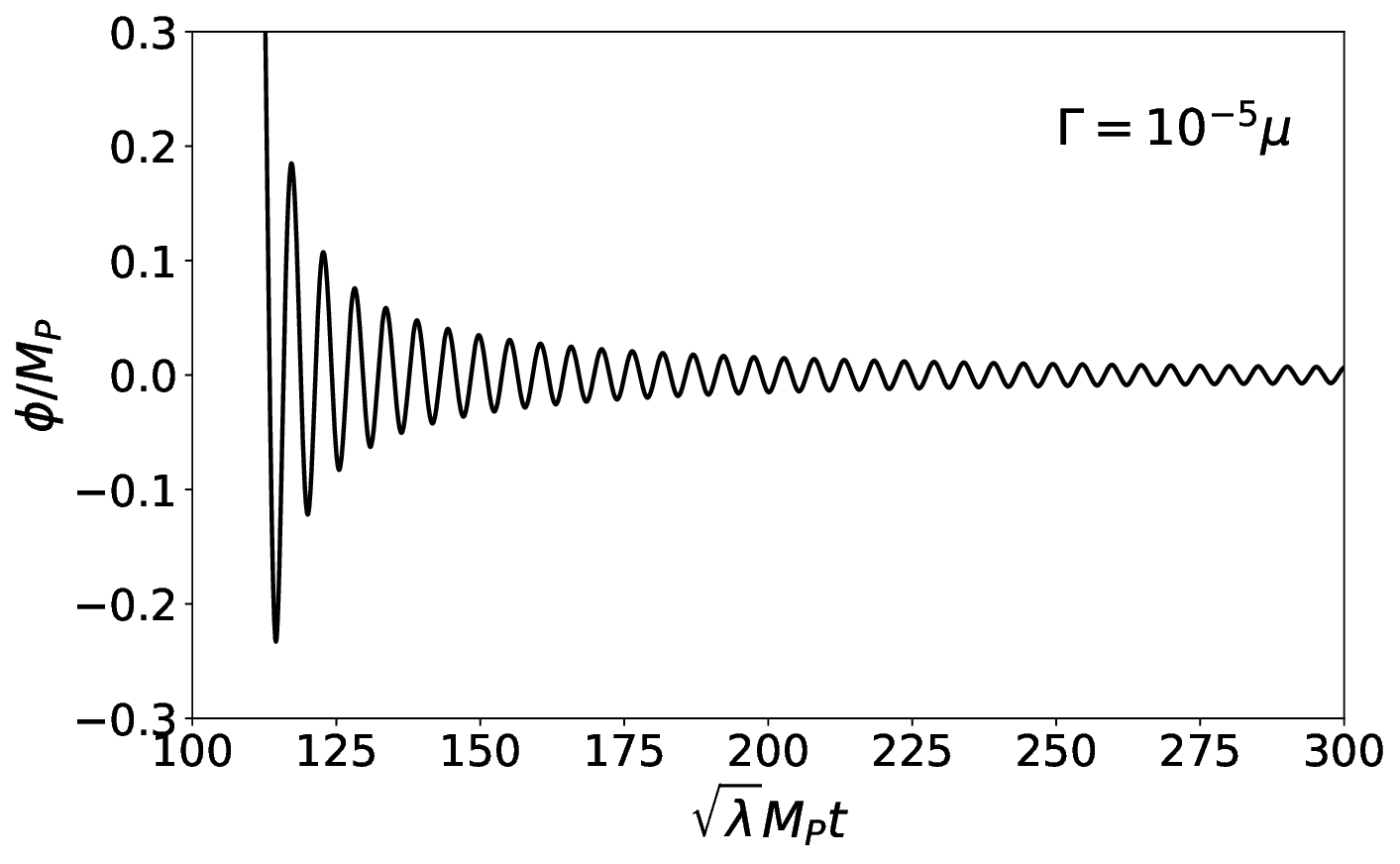}
\caption{\small Time evolution of the scalaron field $\phi(t)$ resulting from numerical integration of the coupled nonlinear dynamical equations \ref{u-tilde}, \ref{rho-tilde} and \ref{phi-tilde}. Left panels: Scalaron field $\phi(t)$ in the inflationary regime ending at $\sqrt{\lambda} M_P t_f=112.0000$ (for $\Gamma=\mu$ and $\Gamma=10^{-5}\mu$). These plots also show some part of the the post-inflationary reheating regime. Right panels: The reheating regime for the cases $\Gamma=\mu$ and $\Gamma=10^{-5}\mu$ showing damped oscillation of the scalaron field, the damping is due both to radiation creation by scalaron decay and Hubble expansion of the Universe.}
\label{fig-1}
\end{figure*}

Starting with a radiation-free universe [$\tilde\rho(0)=0$], we observe slow growth in radiation in the inflationary period. Upon attaining the end of inflation, we carry forward the numerical integration to smoothly enter into the post-inflationary reheating phase. Post-inflation, we observe a significant growth in radiation. The results of the numerical integration for both phases are detailed below.

In Figure~\ref{fig-1}, we display the time evolution of the scalaron field $\phi(t)$ in both regimes, inflation and post-inflationary reheating. With the initial value of $\phi_i=5.4531\, M_P$, our numerical data indicates that the scalaron field diminishes to $\phi_f=0.5819\, M_P$ at the end of inflation (identified by the slow roll parameter $\epsilon$ reaching unity) at the instant $\sqrt \lambda M_P t_f=112.0000$ or $t_f=9.9482\times\,10^{6}\,t_P$ for both $\Gamma=\mu$  and $\Gamma=10^{-5}\mu$. Here $t_P\ (=2.70\times 10^{-43}$ sec) is the reduced Planck time.

For $t>t_f$, that is post-inflation, the scalaron field undergoes a damped oscillation, as illustrated more clearly in the right-hand panels in Figure~\ref{fig-1}.

In Figure~\ref{fig-2}, we display time evolutions of the slow roll parameter $\epsilon(t)$ and the number of e-folds $N(t)$. The slow roll parameter $\epsilon(t)$ remains very small for a prolonged period during inflation, and it attains the value $\epsilon \approx 1$ at $\sqrt{\lambda} M_P t_f = 112.0000$, indicating the end of inflation, for both $\Gamma = \mu$ and $\Gamma = 10^{-5}\mu$. In contrast, the number of e-folds grows more quickly, attaining the same value $N_f = 61.5465$ for both $\Gamma = \mu$ and $\Gamma = 10^{-5}\mu$ at the end of inflation.

Figure~\ref{fig-3} displays the Hubble expansion rate $H(t)$, computed from the Friedmann equation \ref{fried1}, which includes the backreaction from the created radiation energy density $\rho$. The graphs begin with an initial value of $H_i = 0.5706 \sqrt{\lambda} M_P$ and reduce to about half its initial value during inflation, attaining $H_f = 0.2725 \sqrt{\lambda} M_P$ for both $\Gamma = \mu$ and $\Gamma = 10^{-5} \mu$ at the end of inflation. Such slow change in the Hubble rate signifies a quasi-de Sitter phase of the Universe. The Hubble rate exhibits a wavy decline in the ensuing post-inflationary reheating phase, reflecting the effect of damped oscillation of the scalaron field shown in Figure~\ref{fig-1}.

Figure~\ref{fig-4} compares the evolution of the radiation energy density $\rho(t)$ with the energy density $u(t)$ of the scalaron field in both the inflationary and post-inflationary reheating phases. As a result of slow radiation creation during inflation, the radiation energy density $\rho$ gradually increases from an initial value of $\rho_i = 0$, reaching $\rho_f = 1.9635 \times 10^{-11} \lambda M_P^4 = 2.4887 \times 10^{-21} M_P^4$ for $\Gamma = \mu$ and $\rho_f = 1.9635 \times 10^{-16} \lambda M_P^4 = 2.4887 \times 10^{-26} M_P^4$ for $\Gamma = 10^{-5} \mu$ by the end of inflation. Conversely, the energy density $u$ stored in the scalaron field decreases from an initial value of $u_i = 0.9769 \lambda M_P^4 = 1.2382 \times 10^{-10} M_P^4$ to $u_f = 0.2228 \lambda M_P^4 = 2.8240 \times 10^{-11} M_P^4$ for both $\Gamma = \mu$ and $\Gamma = 10^{-5} \mu$ by the end of inflation (without any appreciable difference up to four decimal figures).

\section{Inflation and Reheating: Graceful Exit and Kofman-Yi Condition}
\label{sec5}

Table \ref{tab1} shows the results of numerical integration for various quantities at the end of inflation with respect to the six values of $\Gamma$.  Although the radiation energy density $\rho$ of the Universe grows to different values due to radiation creation by the end of inflation, other quantities do not show appreciable difference at the end of the slow-roll phase of inflation.

\begin{figure*}[t]
  \includegraphics[scale=0.32]{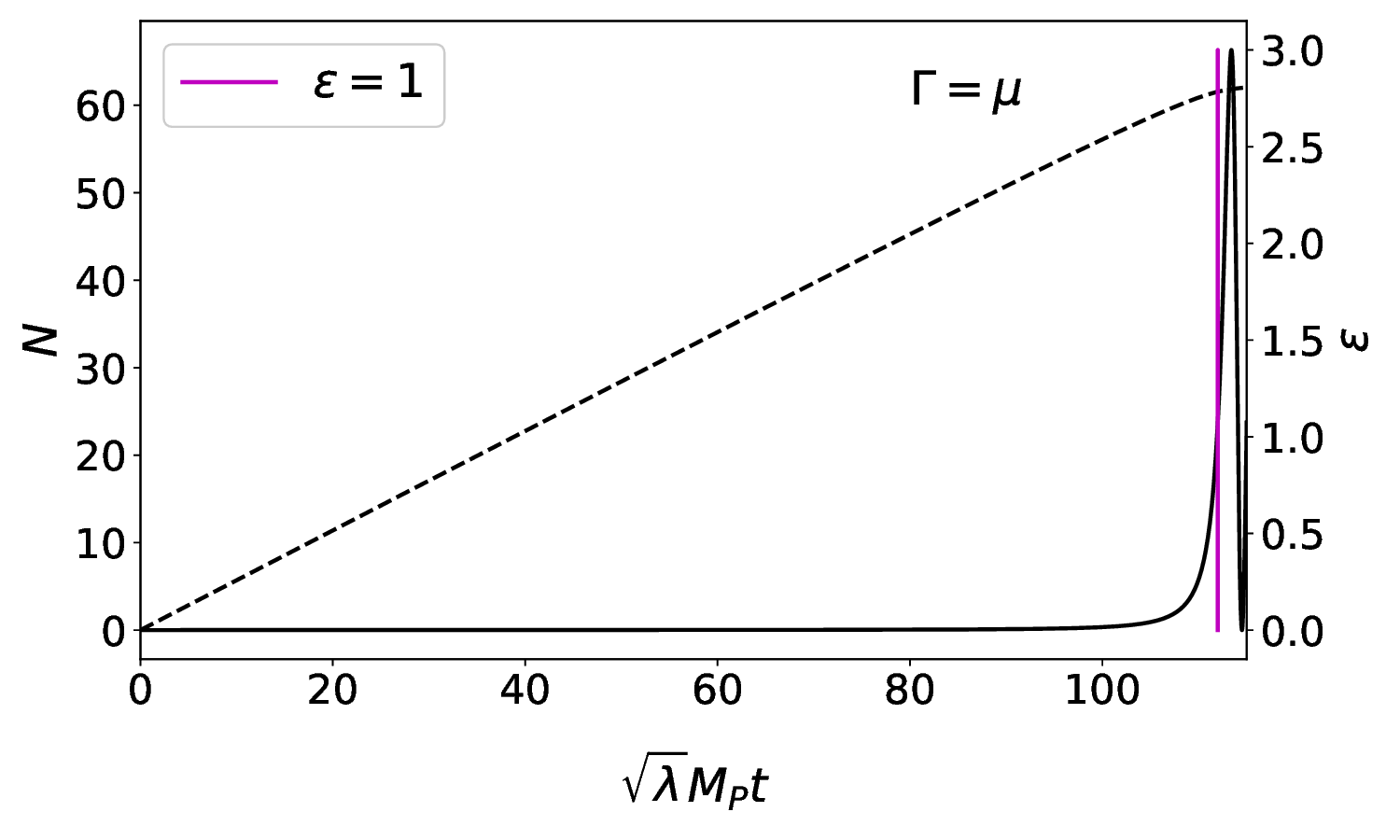}
  \includegraphics[scale=0.32]{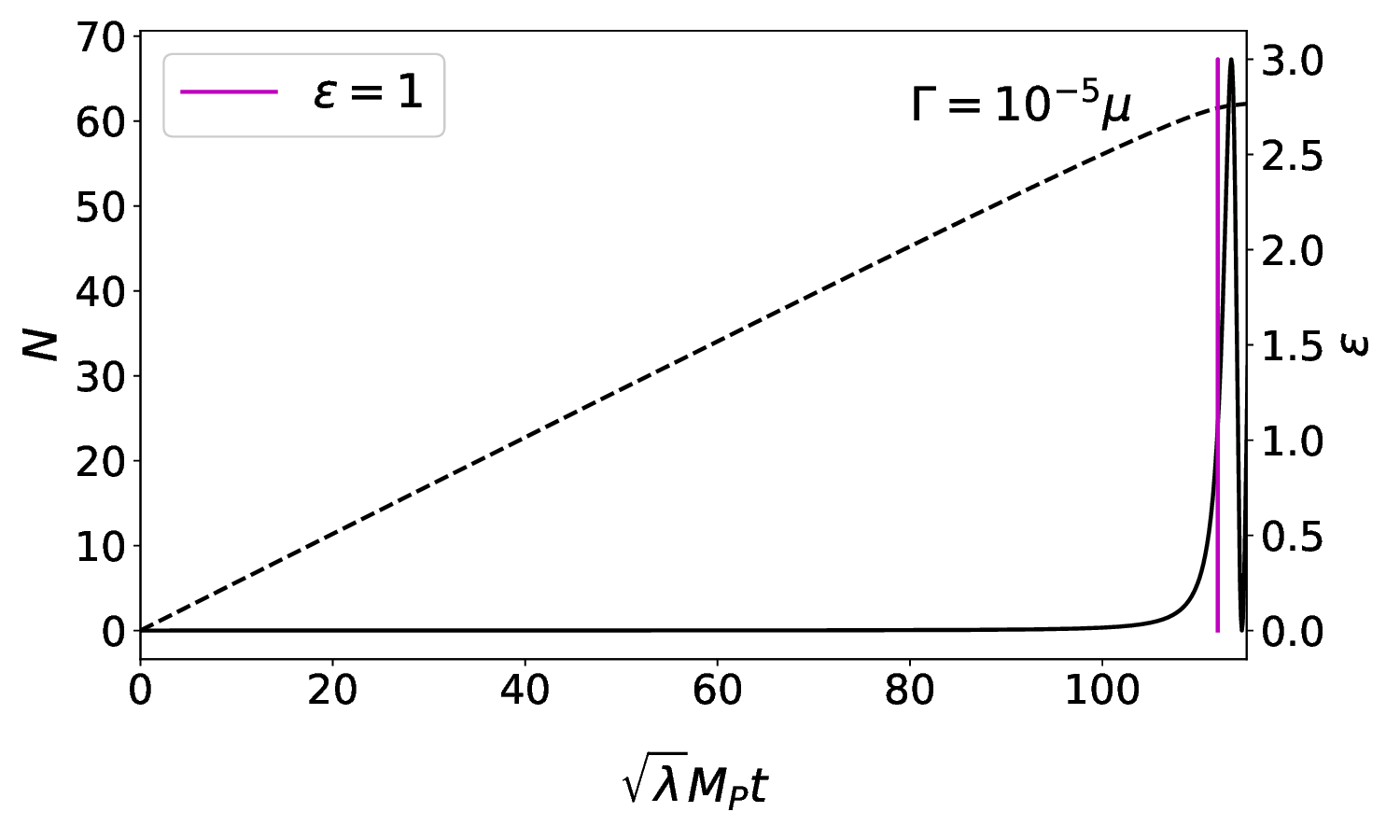}
  \caption{\small The slow roll parameter (solid curves) given by $\epsilon=\frac{3\dot{\phi}^2+4\rho}{\dot{\phi}^2+2V+2\rho}$ and the number of e-folds (dashed curves) given by $N(t)=\int_0^tH dt$. The vertical lines correspond to $\epsilon=1$ marking the end of inflation. For both left panel ($\Gamma=\mu$) and right panel ($\Gamma=10^{-5}\mu$), inflation ends at $\sqrt{\lambda} M_P t_f=112.0000$ with $N_f=61.5465$.}
  \label{fig-2}
\end{figure*}

Importantly, the inequality $u_f\gg\rho_f$ (shown in the table) indicates that the scalaron energy density $u$ overwhelmingly dominates over energy density $\rho$ of the created radiation. This signifies that the inflationary dynamics is principally dictated by the scalaron field and through the usual slow-roll mechanism alone. The created radiation plays a negligible role in the inflationary evolution itself.

Moreover, the Hubble parameter $H_f$ at the end of inflation remains virtually unchanged for different values of $\Gamma$. This demonstrates that the created radiation has practically no impact on the background dynamics during inflation and does not interfere with achieving sufficient e-folds.

As also shown in Table \ref{tab1}, the different cases of $\Gamma$ has no effect on the total number of e-folds, confirming that the scalaron decay into radiation does not compromise the inflationary expansion.

\begin{figure*}[t]
  \includegraphics[scale=0.35]{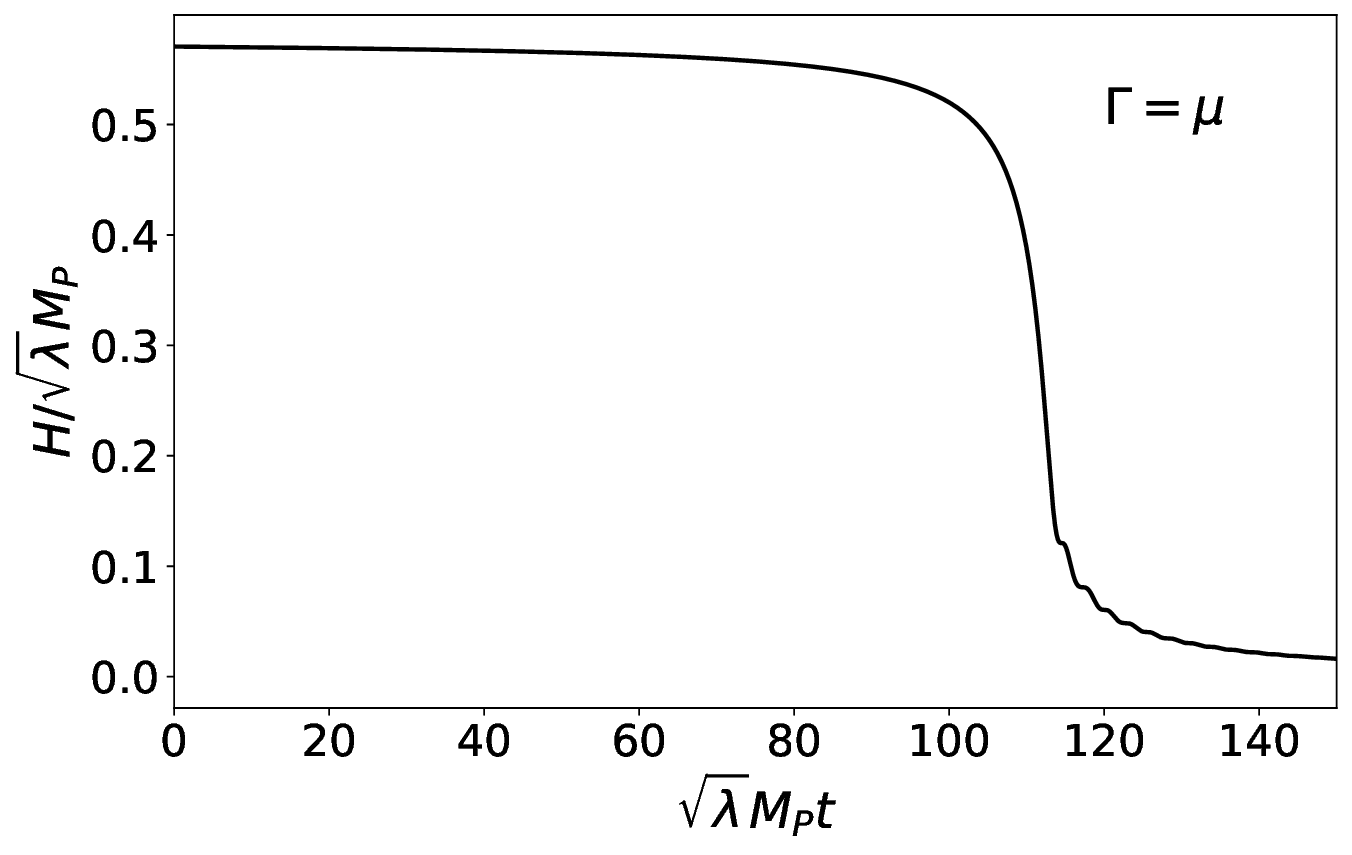}
  \includegraphics[scale=0.35]{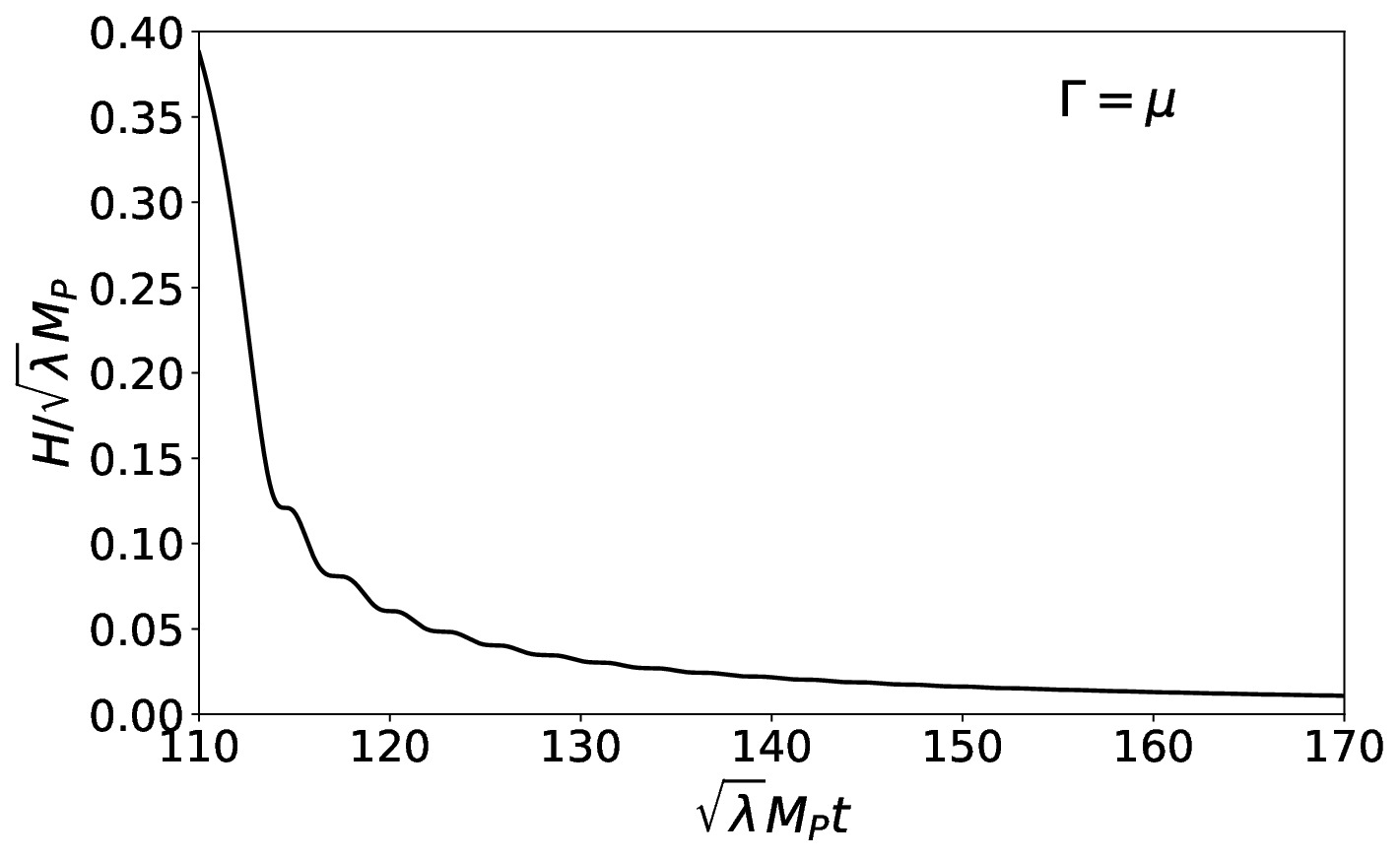}
  \includegraphics[scale=0.35]{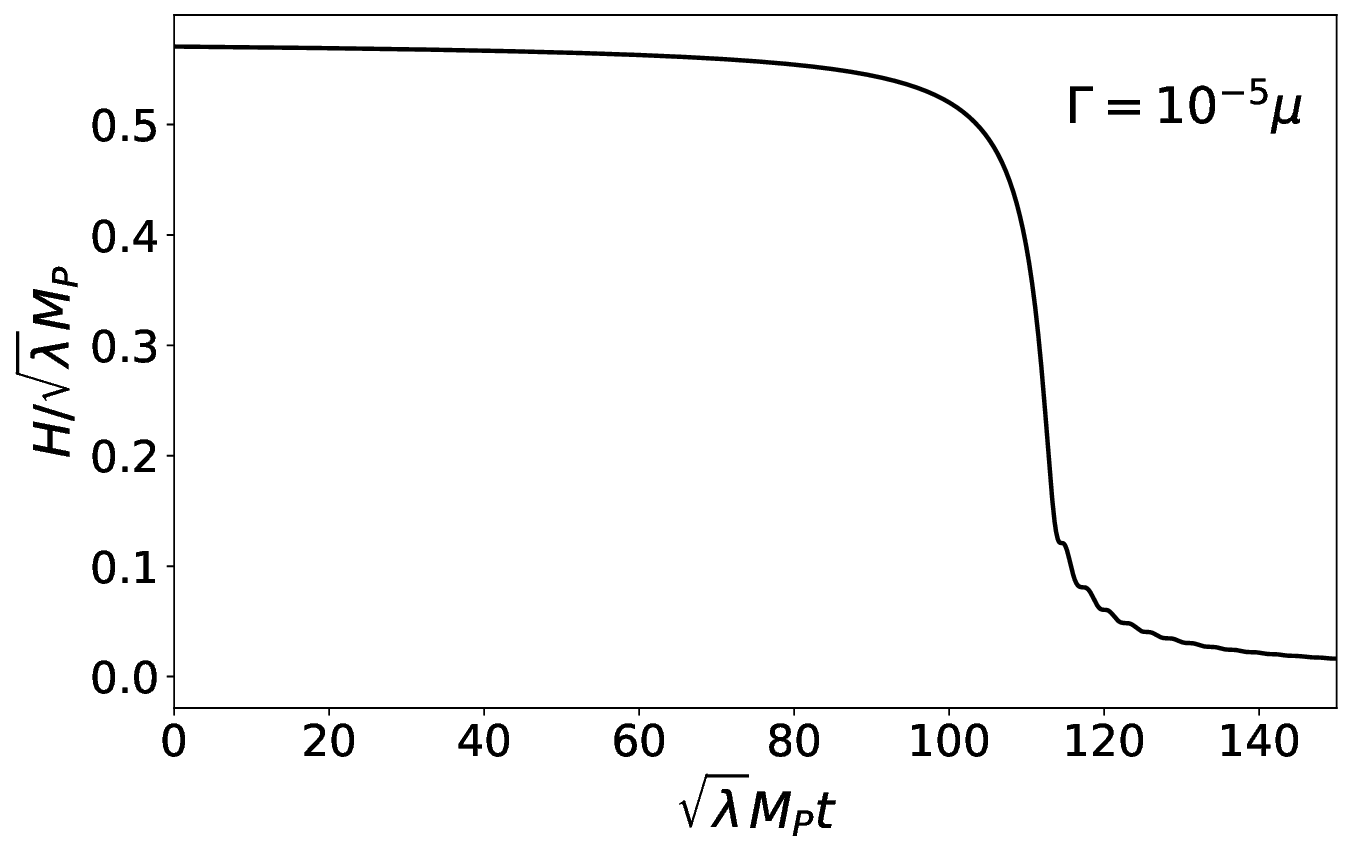}
  \includegraphics[scale=0.35]{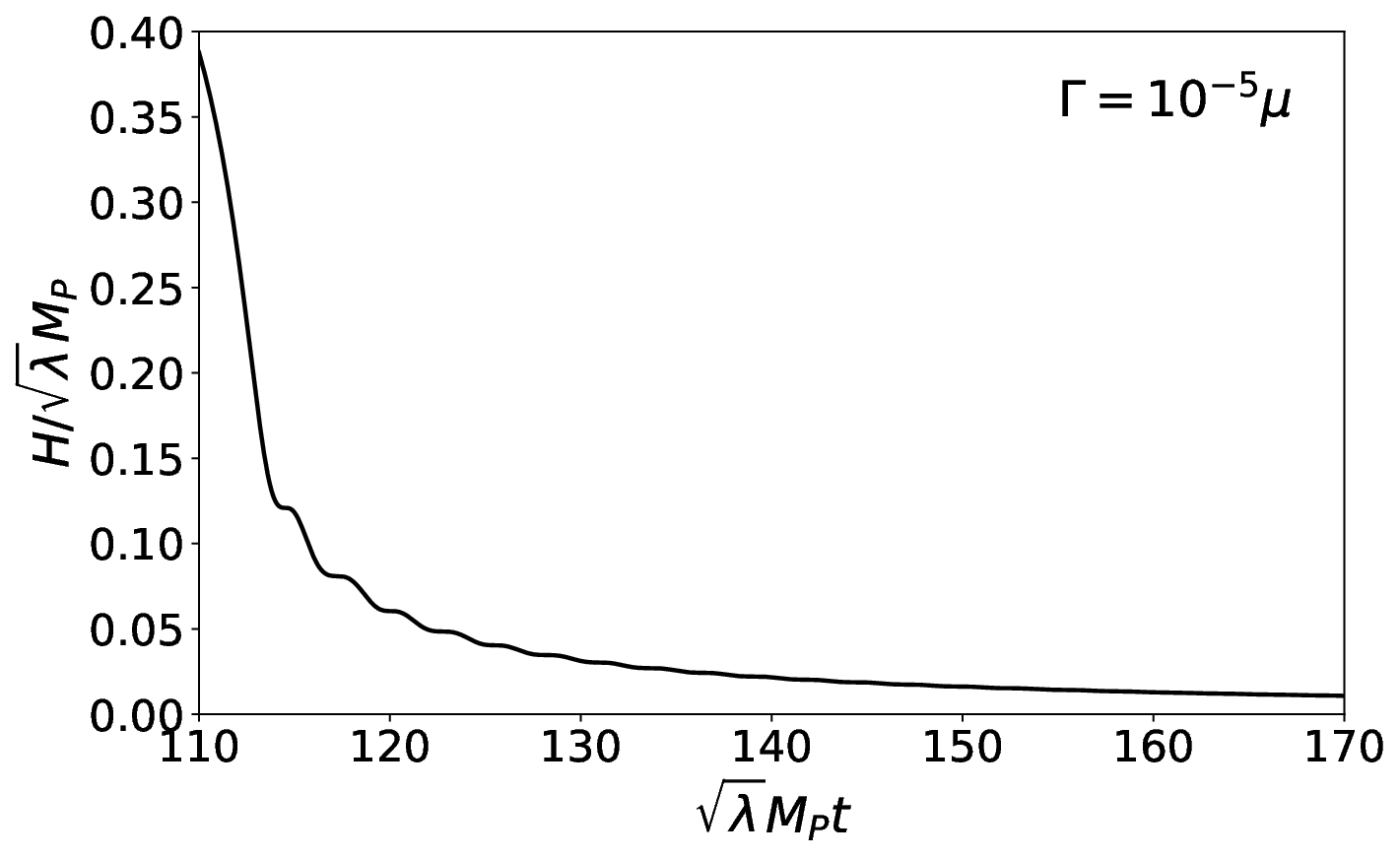}
\caption{\small Hubble expansion rate $H(t)$ with backreaction from scalaron decay given by Friedmann equation \ref{fried1}. Left panels: $H(t)$ in the inflationary regime which ends at $\sqrt{\lambda} M_P t_f=112.0000$ for $\Gamma=\mu$ and $\Gamma=10^{-5}\mu$. Right panels: Hubble expansion rate $H(t)$ after inflation for $\Gamma=\mu$ and $\Gamma=10^{-5}\mu$.}
\label{fig-3}
\end{figure*}

\begin{figure*}[t]
  \includegraphics[scale=0.33]{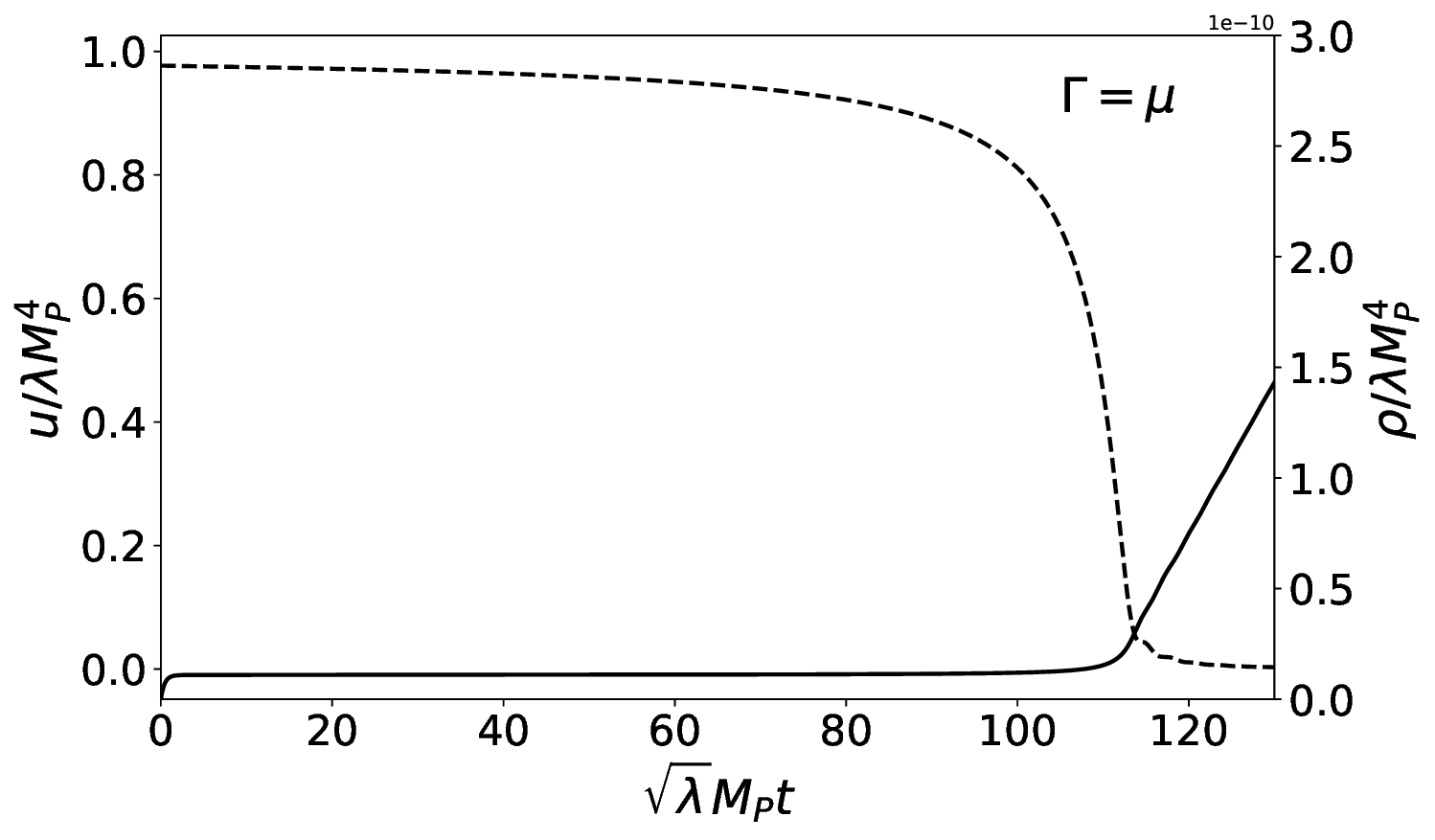}
  \includegraphics[scale=0.33]{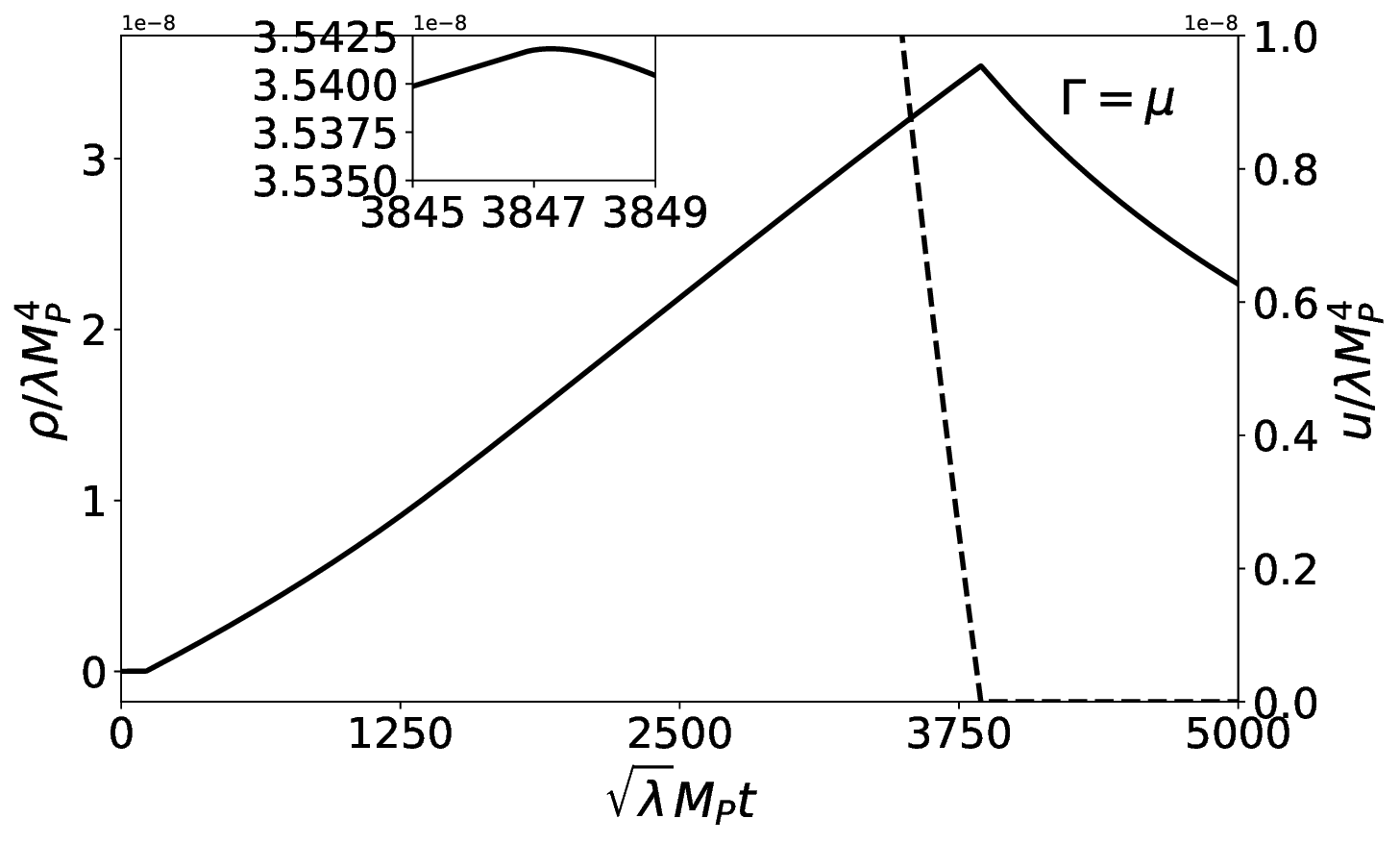}
  \includegraphics[scale=0.33]{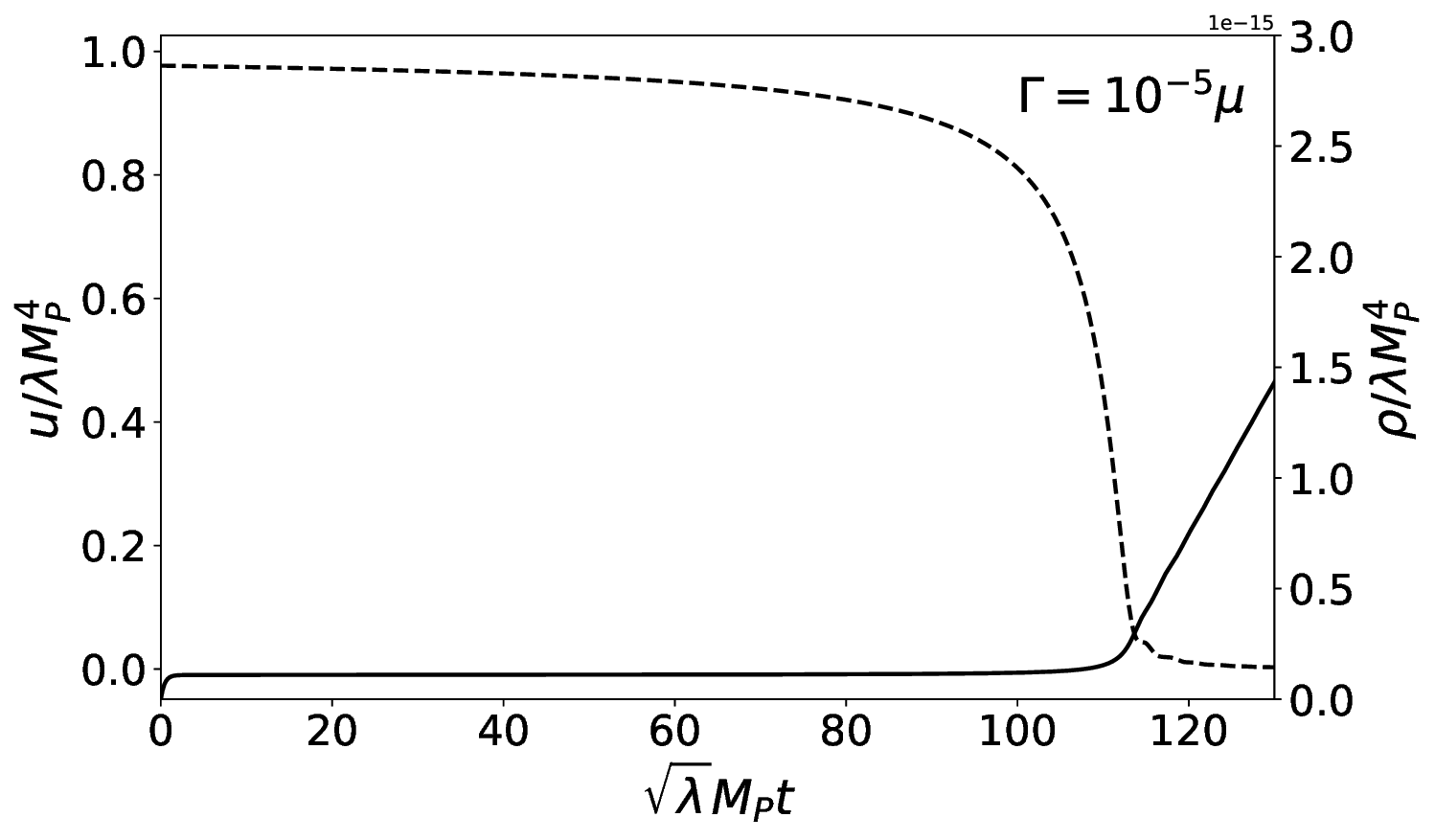}
  \includegraphics[scale=0.33]{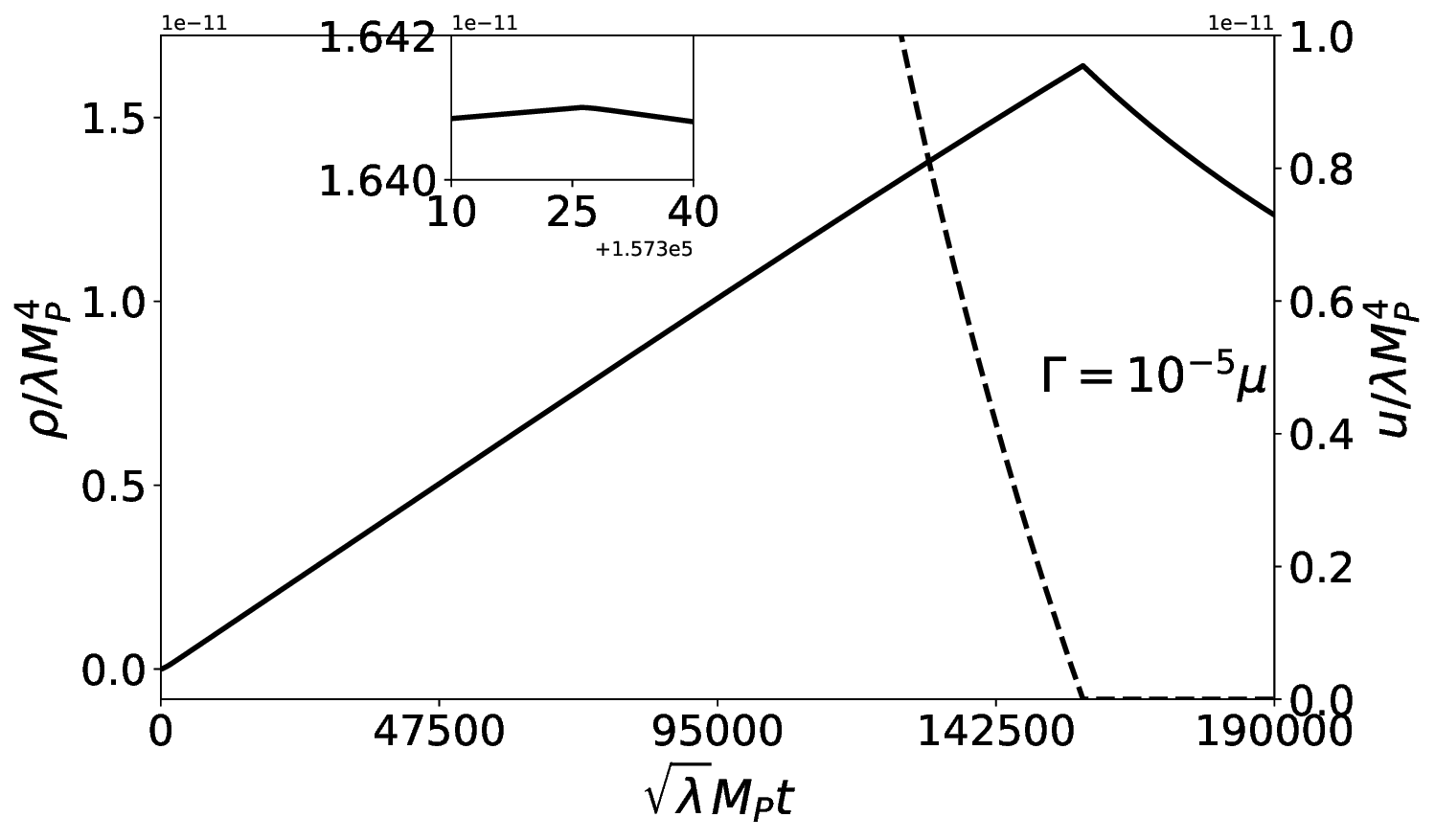}
\caption{\small Growth in energy density $\rho(t)$ (solid curves) and the corresponding depletion in scalaron energy density $u(t)$ (dashed curves). Left panels: For $\Gamma=\mu$ ($\Gamma=10^{-5}\mu$), $\rho$ grows to a value of $\rho_f=1.9635\times 10^{-11}\,\lambda M_P^4$ ($\rho_f=1.9635\times 10^{-16}\,\lambda M_P^4$), and correspondingly $u$ decays to $u_f=0.2228\,\lambda M_P^4$ ($u_f=0.2228\,\lambda M_P^4$) at the end of inflation at $\sqrt{\lambda} M_P t_f=112.0000$ ($\sqrt{\lambda} M_P t_f=112.0000$), at the end of inflation. Right panels: In the post-inflationary reheating phase, the energy density $\rho$ undergoes a substantial growth and attains a peak value of $\rho_r=3.5418\times 10^{-8}\,\lambda M_P^4$ ($\rho_r=1.6410\times 10^{-11}\,\lambda M_P^4$) at $\sqrt{\lambda} M_P t_r=3847.2799$ ($\sqrt{\lambda} M_P t_r=157326.0960$), marking the end of reheating. The insets illustrate smoothness of $\rho(t)$ at the maxima $\rho_r$. The scalaron energy density $u(t)$ undergoes a rapid falloff (dashed curves) in the late stage of reheating and it  falls to a value of $u_r=1.5384\times\,10^{-11}\,\lambda M_P^4$ ($u_r=1.5842\times\,10^{-16}\,\lambda M_P^4$) when $\rho$ attains the peak value $\rho_r$. These values indicate $\rho_r \gg u_r$, making the Universe radiation-dominated, satisfying the Kofman-Yi criterion for successful reheating.}
\label{fig-4}
\end{figure*}

As Figures \ref{fig-1}, \ref{fig-2}, \ref{fig-3}, and \ref{fig-4} illustrate, once the inflatinary period ends at $\tau_f=112$ on completion of $N_f=61.5$ e-folds when the slow-roll parameter attains the value $\epsilon\approx1$, there occurs a smooth entry into the post-inflatinary regime marked by the onset of an oscillatory scalaron field $\phi(t)$ with the radiation component $\rho(t)$ acquiring a significant growth rate.

This behavior contitutes a {\em graceful exit\/} from the inflationary regime and a {\em smooth entry} into the next phase, the reheating phase of the Universe. As the right panels in Figure \ref{fig-4} illustrate, the Universe ultimately becomes radiation-dominated and arrives at the condition $\rho_r\gg  u_r$ at the time when the radiation energy density $\rho$ attains a peak while the scalaron energy density exhibits a sharp falloff so that the scalaron component becomes negligibly small. This is the right condition consistent with the {\em Kofman-Yi criterion}~\ref{koff} that requires a hot, radiation-dominated universe so that subsequent big-bang cosmological processes can take place.

The above qualitative analysis of the post-inflationary reheating regime ($t>t_f$) can be arrived at more precisely through the following paragraphs, written on the basis of numerical data.

During an extended period of time in the post-inflationary reheating phase ($\tau>\tau_f=112.0000$), the energy density $u$ available in the scalaron field remains sufficient despite the ongoing scalaron-energy drainage to radiation, consistently meeting the sufficiency condition $u>\frac{8\varepsilon ^4}{3\sqrt{3}}$. Our numerical data indicates that $u$ remains sufficient until $\sqrt{\lambda} M_P t_* = 3846.7899$ for $\Gamma = \mu$ and until $\sqrt{\lambda} M_P t_* = 99729.9945$ for $\Gamma = 10^{-5} \mu$. During this time, the sink term $-\tilde{q}$ remains constant. After the time $t_*$, the energy density $u$ remaining in the scalaron field becomes insufficient, and the sink term transitions to $-\Gamma u$, causing $u$ to deplete rapidly. Then, at $\sqrt{\lambda} M_P t_r = 3847.2799$ for $\Gamma = \mu$ and at $\sqrt{\lambda} M_P t_r = 157326.0960$ for $\Gamma = 10^{-5} \mu$, the radiation energy density $\rho$ reaches its peak (Figure~\ref{fig-4}), that may be identified as the end of reheating. Following this peak, the Universe cools as $\rho$ continues to decrease.

From our numerical data in the post-inflationary reheating regime, we find that the radiation energy density $\rho(t)$ attains its peak value $\rho_r$ at time $t_r$, these values being
 $\rho_r=3.5418\times10^{-8}\,\lambda M_P^4=4.4892\times10^{-18}\,M_P^4$ at $\sqrt \lambda M_P t_r=3847.2799$ for $\Gamma=\mu$, and 
 $\rho_r=1.6410\times10^{-11}\,\lambda M_P^4=2.0800\times10^{-21}\,M_P^4$ at $\sqrt \lambda M_P t_r=157326.0960$ for $\Gamma=10^{-5}\mu$. 
The insets in Figure~\ref{fig-4} highlight the smoothness of the curve $\rho(t)$ at the maxima $\rho_r$. 

In contrast, the dashed curves in the right-hand panel of Figure~\ref{fig-4} illustrate rapid falloff in the scalaron energy density $u(t)$. At time $t_r$, the energy density $u$ drops to a value of $u_r$, given by    $u_r=1.5384\times\,10^{-11}\,\lambda M_P^4=1.9499\times 10^{-21}M_P^4$ for $\Gamma=\mu$, and $u_r=1.5842\times\,10^{-16}\,\lambda M_P^4=2.0080\times 10^{-26}M_P^4$ for $\Gamma=10^{-5}\mu$.

Importantly, the above numerical results implying $\rho_r\gg u_r$ indicate that the Universe becomes radiation-dominated satisfying the {\em Kofman-Yi criterion} \ref{koff}.

We can relate the maxima $\rho_r$ of the created radiation energy density with the reheating temperature $T_r$ via the relation
\begin{equation}
\rho_r = \frac{\pi^2}{30} g_* T_r^4,
\end{equation}
where $g_*$ represents the total number of degrees of freedom in the Standard Model of particle physics. At energies exceeding 1 TeV, all Standard Model particles are relativistic, which gives $g_* = 106.75$ \cite{d2017universe}.

Table \ref{tab2} summarizes the results of the numerical integration for various quantities at the end of reheating, including the reheating temperature $T_r$, determined from the maximum radiation energy density $\rho_r$, for six different values of $\Gamma/\mu$ ranging from unity to $10^{-5}$. In all cases, the radiation energy density significantly exceeds the residual scalaron energy density, i.e., $\rho_r \gg u_r$, at the end of reheating. This demonstrates that the Universe has entered a radiation-dominated phase satisfying the {\em Kofman--Yi criterion}.

In summary, the dynamical evolution provides a smooth {\em graceful exit} from inflation into the reheating regime where radiation is produced rapidly at the expense of the scalaron energy. As reheating proceeds, the scalaron energy is progressively depleted, while the radiation energy density grows and eventually dominates the cosmic energy. The Universe thus evolves into a hot, radiation-dominated phase, with the scalaron contribution becoming negligible, thereby satisfying the {\em Kofman--Yi criterion}.

\begin{table*}
\caption{\small Results at the end of inflation with different scalaron decay rates $\Gamma$ by numerical integration of the coupled nonlinear dynamical equations \ref{u-tilde}, \ref{rho-tilde} and \ref{phi-tilde}. Although the energy density $\rho$ of the Universe grows to different values due to radiation creation from spontaneous scalaron deccay during inflation, other quantities do not show appreciable difference at the end of the slow-roll phase of inflation.}
\begin{center}
    \begin{tabular} {  c | c | c | c | c | c | c | c  }
      \hline\hline
    \multicolumn{8} { c  }{End of Inflation}\\
    \hline
     $\Gamma/\mu$ & $\tau_f$ & $\epsilon_f$ & $N_f$ & $\tilde \phi_f$  & $\tilde H_f$ & $\tilde u_f$ & $\tilde \rho_f$ \\
    \hline
    1             & 112.0000 & 1.0740 & 61.5465 & 0.5819  & 0.2725 & 0.2228 & $1.9635\times 10^{-11}$ \\
    \hline
    $10^{-1}$     & 112.0000 & 1.0740 & 61.5465 & 0.5819  & 0.2725  & 0.2228 & $1.9635\times 10^{-12}$ \\
    \hline
    $10^{-2}$    & 112.0000 & 1.0740 & 61.5465 & 0.5819   & 0.2725 & 0.2228 & $1.9635\times 10^{-13}$ \\
    \hline
    $10^{-3}$    & 112.0000 & 1.0740 & 61.5465 & 0.5819   & 0.2725 & 0.2228 & $1.9635\times 10^{-14}$ \\
    \hline
    $10^{-4}$    & 112.0000 & 1.0740 & 61.5465 & 0.5819   & 0.2725 & 0.2228  & $1.9635\times 10^{-15}$ \\
    \hline
    $10^{-5}$    & 112.0000 & 1.0740 & 61.5465 & 0.5819   & 0.2725 & 0.2228 & $1.9635\times 10^{-16}$ \\
    \hline\hline
    \end{tabular}
\end{center}
      \label{tab1}
\end{table*}

\section{Discussion and Conclusion}
\label{discussion} 

In this study, we employed the Starobinsky scenario to examine the behavior of the primordial Universe in the inflationary and the post-inflationary reheating phases. Given the {\em spontaneous} nature of the scalaron decay, we were prompted to explore radiation creation from scalaron decay during both phases, the slow-roll phase of inflation and in the post-inflationary reheating phase.

We specifically focused on the decay channel $\phi \rightarrow \chi \chi$, which is significantly more efficient than $\phi \phi \rightarrow \chi \chi$, the latter being Planck-suppressed as evident from equation (\ref{mat2}).

To model the associated production of radiation, we used a physically motivated expression for the rate of energy gain in the radiation field, represented by a source term in the evolution equation for the radiation energy density $\rho(t)$. By enforcing exact energy conservation, we ensured that the energy gained by the radiation field precisely compensates for the energy lost by the scalaron. This yielded, in the expandng background driven by the Friedmann equation, a coupled set of nonlinear differential equations for the radiation energy density $\rho(t)$ and the scalaron energy density $u(t)$, with the source and sink terms being equal in magnitude but opposite in sign. The backreaction of the created radiation density was incorporated into the Friedmann equation, allowing us to track the dynamics of the universe self-consistently.

We numerically integrated this set of coupled nonlinear differential equations, initialized with a radiation-free state, from the onset of inflation through to the reheating regime. The initial conditions in the inflatinary period were chosen to satisfy slow-roll criteria so as to produce approximately 60 e-folds of inflation. At the point where the slow-roll parameter $\epsilon$ reached unity, marking the end of inflation, the integration was continued smoothly into the post-inflationary reheating regime.

Our numerical results show that, although radiation is continuously generated during inflation, its energy density $\rho$ remains subdominant compared to the scalaron energy density $u$. For example, $\tilde\rho$ increases from zero to  $\sim10^{-11}$ for $\Gamma = \mu$ and to   $\sim10^{-16}$ for $\Gamma = 10^{-5}\mu$, while $\tilde u$ decreases from $\sim1$ to $\sim0.2$ at the end of inflation, thus maintaining scalaron dominance throughout inflation.  The Hubble expansion rate, in turn, reduces by roughly a factor of two, indicating that the Universe experienced a quasi-de Sitter expansion during inflation.

We performed simulations for multiple decay rates, $\Gamma/\mu = 1, 10^{-1}, 10^{-2}, 10^{-3}, 10^{-4}, 10^{-5}$. As shown in Table~\ref{tab1}, while the radiation energy density $\rho$ grows to different values with respect to $\Gamma$, other quantities remain largely unaffected due to continued dominance of the scalaron during inflation. This confirms that standard slow-roll inflation persists despite ongoing radiation production, and the scalaron field remains the principal driver of cosmic inflation.

Following the inflationary phase, the scalaron smoothly transitions into a regime of damped oscillations, marking the onset of reheating. During this stage, the scalaron continuously transfers its energy to the radiation sector through its decay, causing the radiation energy density $\rho$ to grow substantially even for small decay rates, while the scalaron energy density $u$ steadily decreases.
The coupled dynamical evolution of the scalaron and radiation components thus realizes a smooth {\em graceful exit} from inflation into the reheating phase, while consistently accounting for the transfer of energy between the two sectors. This continuous energy transfer builds up a substantial radiation ddensity even for small decay rates, providing the conditions necessary for the subsequent radiation-dominated evolution.

\begin{table*}
\caption{\small Results at the end of reheating with different scalaron decay rates $\Gamma$ by numerical integration of the coupled nonlinear dynamical equations \ref{u-tilde}, \ref{rho-tilde} and \ref{phi-tilde}. The end of reheating is identified by the radiation energy density $\rho$ attaining its peak value $\rho_r$.}
    \begin{center}
    \begin{tabular} {  c | c | c | c | c | c  }
      \hline\hline
   \multicolumn{6} { c }{End of Reheating}\\
    \hline
     $\Gamma/\mu$ & $\tau_r$ & $\tilde u_r$ & $\tilde H_r$ & $\tilde \rho_r$ & $T_r$/GeV\\
    \hline
          $1$ & $3847.2799$  & $1.5384\times 10^{-11}$  & $1.0868\times 10^{-4}$  & $3.5418\times 10^{-8}$  & $4.5380\times 10^{13}$\\
    \hline
    $10^{-1}$ & $7751.1798$  & $1.5361\times 10^{-12}$ & $5.0399\times 10^{-5}$  & $7.6187\times 10^{-9}$  & $3.0905\times 10^{13}$\\
    \hline
    $10^{-2}$ & $16156.6996$ & $1.5221\times 10^{-13}$ & $2.3389\times 10^{-5}$  & $1.6410\times 10^{-9}$  & $2.1054\times 10^{13}$\\
    \hline
    $10^{-3}$ & $34264.6991$ & $1.5543\times 10^{-14}$ & $1.0856\times 10^{-5}$  & $3.5354\times 10^{-10}$ & $1.4344\times 10^{13}$\\
    \hline
    $10^{-4}$ & $73276.8987$ & $1.5990\times 10^{-15}$ & $5.0388\times 10^{-6}$  & $7.6168\times 10^{-11}$ & $9.6253\times 10^{12}$\\
    \hline
    $10^{-5}$ & $157326.0960$ & $1.5842\times 10^{-16}$ & $2.3388\times 10^{-6}$ & $1.6410\times 10^{-11}$ & $6.6579\times 10^{12}$\\
    \hline
    \hline
    \end{tabular}
    \end{center}
    \label{tab2}
\end{table*}

Notably, the scalaron energy density remains sufficient to sustain efficient radiation production until $\sqrt{\lambda} M_P t_* = 3846.7899$ for $\Gamma = \mu$ and until $\sqrt{\lambda} M_P t_* = 99729.9945$ for $\Gamma = 10^{-5} \mu$. Thereafter, the radiation energy density $\rho$ reaches its peak at $\sqrt{\lambda} M_P t_r = 3847.2799$ for $\Gamma = \mu$ and at $\sqrt{\lambda} M_P t_r = 157326.0960$ for $\Gamma = 10^{-5} \mu$.

The peak radiation energy densities are found to be $\rho_r = 1.4894 \times 10^{56}$ GeV$^4$ for $\Gamma = \mu$ and $\rho_r = 6.9009 \times 10^{52}$ GeV$^4$ for $\Gamma = 10^{-5} \mu$, corresponding to reheating temperatures of $T_r = 4.5380 \times 10^{13}$ GeV and $T_r = 6.6579 \times 10^{12}$ GeV, respectively.

At the peak radiation energy density $\rho_r$, the corresponding scalaron energy density $u_r$ is found to be substantially smaller, with $\rho_r \gg u_r$ for all values of the decay constant considered, as shown in Table~\ref{tab2}. Since the Universe initially evolves from a scalaron-dominated, radiation-free state, this strong hierarchy demonstrates that a large fraction of the scalaron energy is transferred to radiation during reheating. The resulting state is therefore overwhelmingly radiation dominated and can provide the appropriate high-temperature environment required for the subsequent hot Big-Bang evolution. This result satisfies the Kofman--Yi criterion \cite{kofman_PRD_2005} for successful reheating.

In our framework, this condition is realized naturally throughout the range of scalaron decay rates $\Gamma$ considered. The time at which $\rho_r$ reaches its maximum thus provides a natural marker for the completion of the energy-transfer process and the onset of the radiation-dominated phase. The condition $\rho_r \gg u_r$ at this stage confirms that the residual scalaron contribution is negligible and that the energy budget of the Universe has been effectively transferred to radiation. Consequently, the reheating dynamics in our model provide a smooth transition from the inflationary scalaron-dominated regime to a hot, radiation-dominated Universe, establishing the appropriate conditions for subsequent Big-Bang cosmological evolution.

In conclusion, our analysis shows that the Starobinsky model provides a coherent dynamical framework for graceful exit from inflation to reheating. The coupled evolution of the scalaron and radiation components in the expanding background leads to efficient energy transfer and satisfies the Kofman-Yi criterion for successful reheating, resulting in a radiation-dominated, high-temperature state suitable for the subsequent hot Big-Bang evolution. The dynamical evolution aligns well with theoretical expectations and support viability of the Starobinky model as a compelling model of the early Universe.

\section*{Acknowledgements}

Bhargabi Saha is indebted to Professor A. A. Starobinsky for motivating discussions in a conference held at Indian Institute of Science Education and Research (IISER), Mohali. She is supported by the Ministry of Human Resource Development (MHRD), Government of India, through a Research Fellowship. The Authors would like to thank the Indian Institute of Technology Guwahati for providing access to computing facilities.


\begin{thebibliography}{10}
\expandafter\ifx\csname url\endcsname\relax
  \def\url#1{\texttt{#1}}\fi
\expandafter\ifx\csname urlprefix\endcsname\relax\def\urlprefix{URL }\fi
\expandafter\ifx\csname href\endcsname\relax
  \def\href#1#2{#2} \def\path#1{#1}\fi

\bibitem{DAVIES1977}
P.~Davies,
  \href{https://www.sciencedirect.com/science/article/pii/0370269377905044}{Singularity
  avoidance and quantum conformal anomalies}, Physics Letters B 68~(4) (1977)
  402--404.
\newblock \href {https://doi.org/https://doi.org/10.1016/0370-2693(77)90504-4}
  {\path{doi:https://doi.org/10.1016/0370-2693(77)90504-4}}.
\newline\urlprefix\url{https://www.sciencedirect.com/science/article/pii/0370269377905044}

\bibitem{DAVIES-EMT}
P.~Davies, S.~Fulling, S.~Christensen, T.~Bunch,
  \href{https://www.sciencedirect.com/science/article/pii/0003491677901671}{Energy-momentum
  tensor of a massless scalar quantum field in a robertson-walker universe},
  Annals of Physics 109~(1) (1977) 108--142.
\newblock \href {https://doi.org/https://doi.org/10.1016/0003-4916(77)90167-1}
  {\path{doi:https://doi.org/10.1016/0003-4916(77)90167-1}}.
\newline\urlprefix\url{https://www.sciencedirect.com/science/article/pii/0003491677901671}

\bibitem{bunch1977stress}
T.~Bunch, P.~Davies, Stress tensor and conformal anomalies for massless fields
  in a robertson--walker universe, Proceedings of the Royal Society of London.
  A. Mathematical and Physical Sciences 356~(1687) (1977) 569--574.

\bibitem{Birrell_Davies_1982}
N.~D. Birrell, P.~C.~W. Davies, Quantum Fields in Curved Space, Cambridge
  Monographs on Mathematical Physics, Cambridge University Press, U.K., 1982.

\bibitem{STAROBINSKY1980}
A.~Starobinsky,
  \href{https://www.sciencedirect.com/science/article/pii/037026938090670X}{A
  new type of isotropic cosmological models without singularity}, Physics
  Letters B 91~(1) (1980) 99--102.
\newblock \href {https://doi.org/https://doi.org/10.1016/0370-2693(80)90670-X}
  {\path{doi:https://doi.org/10.1016/0370-2693(80)90670-X}}.
\newline\urlprefix\url{https://www.sciencedirect.com/science/article/pii/037026938090670X}

\bibitem{guth1981inflationary}
A.~H. Guth,
  \href{https://link.aps.org/doi/10.1103/PhysRevD.23.347}{Inflationary
  universe: A possible solution to the horizon and flatness problems}, Phys.
  Rev. D 23 (1981) 347--356.
\newblock \href {https://doi.org/10.1103/PhysRevD.23.347}
  {\path{doi:10.1103/PhysRevD.23.347}}.
\newline\urlprefix\url{https://link.aps.org/doi/10.1103/PhysRevD.23.347}

\bibitem{Sato:1980yn}
K.~Sato, {First Order Phase Transition of a Vacuum and Expansion of the
  Universe}, Mon. Not. Roy. Astron. Soc. 195 (1981) 467--479.

\bibitem{Linde-inflation1982}
A.~D. Linde, {A New Inflationary Universe Scenario: A Possible Solution of the
  Horizon, Flatness, Homogeneity, Isotropy and Primordial Monopole Problems},
  Phys. Lett. B 108 (1982) 389--393.
\newblock \href {https://doi.org/10.1016/0370-2693(82)91219-9}
  {\path{doi:10.1016/0370-2693(82)91219-9}}.

\bibitem{Albrecht1982A}
A.~Albrecht, P.~J. Steinhardt,
  \href{https://link.aps.org/doi/10.1103/PhysRevLett.48.1220}{Cosmology for
  grand unified theories with radiatively induced symmetry breaking}, Phys.
  Rev. Lett. 48 (1982) 1220--1223.
\newblock \href {https://doi.org/10.1103/PhysRevLett.48.1220}
  {\path{doi:10.1103/PhysRevLett.48.1220}}.
\newline\urlprefix\url{https://link.aps.org/doi/10.1103/PhysRevLett.48.1220}

\bibitem{linde1984inflationary}
A.~D. Linde, \href{https://dx.doi.org/10.1088/0034-4885/47/8/002}{The
  inflationary universe}, Reports on Progress in Physics 47~(8) (1984) 925.
\newblock \href {https://doi.org/10.1088/0034-4885/47/8/002}
  {\path{doi:10.1088/0034-4885/47/8/002}}.
\newline\urlprefix\url{https://dx.doi.org/10.1088/0034-4885/47/8/002}

\bibitem{linde1983chaotic}
A.~D. Linde, {Chaotic Inflation}, Phys. Lett. B 129 (1983) 177--181.
\newblock \href {https://doi.org/10.1016/0370-2693(83)90837-7}
  {\path{doi:10.1016/0370-2693(83)90837-7}}.

\bibitem{dolgov1982baryon}
A.~D. Dolgov, A.~D. Linde, {Baryon Asymmetry in Inflationary Universe}, Phys.
  Lett. B 116 (1982) 329.
\newblock \href {https://doi.org/10.1016/0370-2693(82)90292-1}
  {\path{doi:10.1016/0370-2693(82)90292-1}}.

\bibitem{linde-particle}
A.~D. Linde, Particle Physics and Inflationary Cosmology, Harwood, Switzerland,
  1990.

\bibitem{linde1995inflation}
A.~Linde,
  \href{https://www.sciencedirect.com/science/article/pii/037026939500370Z}{Inflation
  with variable $\omega$}, Physics Letters B 351~(1) (1995) 99--104.
\newblock \href {https://doi.org/10.1016/0370-2693(95)00370-Z}
  {\path{doi:10.1016/0370-2693(95)00370-Z}}.
\newline\urlprefix\url{https://www.sciencedirect.com/science/article/pii/037026939500370Z}

\bibitem{linde1988chaotic}
A.~D. Linde, {Chaotic Inflation With Constrained Fields}, Phys. Lett. B 202
  (1988) 194--197.
\newblock \href {https://doi.org/10.1016/0370-2693(88)90006-8}
  {\path{doi:10.1016/0370-2693(88)90006-8}}.

\bibitem{linde1985initial}
A.~D. Linde, {Initial conditions for inflation}, Phys. Lett. B 162 (1985)
  281--286.
\newblock \href {https://doi.org/10.1016/0370-2693(85)90923-2}
  {\path{doi:10.1016/0370-2693(85)90923-2}}.

\bibitem{abbott1982particle}
L.~F. Abbott, E.~Farhi, M.~B. Wise, Particle production in the new inflationary
  cosmology, Physics Letters B 117~(1-2) (1982) 29--33.

\bibitem{traschen1990}
J.~H. Traschen, R.~H. Brandenberger,
  \href{https://link.aps.org/doi/10.1103/PhysRevD.42.2491}{Particle production
  during out-of-equilibrium phase transitions}, Phys. Rev. D 42 (1990)
  2491--2504.
\newblock \href {https://doi.org/10.1103/PhysRevD.42.2491}
  {\path{doi:10.1103/PhysRevD.42.2491}}.
\newline\urlprefix\url{https://link.aps.org/doi/10.1103/PhysRevD.42.2491}

\bibitem{kofman1994reheating}
L.~Kofman, A.~Linde, A.~A. Starobinsky,
  \href{https://link.aps.org/doi/10.1103/PhysRevLett.73.3195}{Reheating after
  inflation}, Phys. Rev. Lett. 73 (1994) 3195--3198.
\newblock \href {https://doi.org/10.1103/PhysRevLett.73.3195}
  {\path{doi:10.1103/PhysRevLett.73.3195}}.
\newline\urlprefix\url{https://link.aps.org/doi/10.1103/PhysRevLett.73.3195}

\bibitem{shtanov1995universe}
Y.~Shtanov, J.~Traschen, R.~Brandenberger,
  \href{https://link.aps.org/doi/10.1103/PhysRevD.51.5438}{Universe reheating
  after inflation}, Phys. Rev. D 51 (1995) 5438--5455.
\newblock \href {https://doi.org/10.1103/PhysRevD.51.5438}
  {\path{doi:10.1103/PhysRevD.51.5438}}.
\newline\urlprefix\url{https://link.aps.org/doi/10.1103/PhysRevD.51.5438}

\bibitem{kofman1996origin}
L.~A. Kofman, {The Origin of matter in the universe: Reheating after
  inflation}, 1996.
\newblock \href {http://arxiv.org/abs/astro-ph/9605155}
  {\path{arXiv:astro-ph/9605155}}.

\bibitem{kofman1998reheating}
L.~Kofman, {Reheating and preheating after inflation}, in: {3rd RESCEU
  International Symposium on Particle Cosmology}, 1997, pp. 1--8.
\newblock \href {http://arxiv.org/abs/hep-ph/9802285}
  {\path{arXiv:hep-ph/9802285}}.

\bibitem{kofman1997towards}
L.~Kofman, A.~Linde, A.~A. Starobinsky,
  \href{https://link.aps.org/doi/10.1103/PhysRevD.56.3258}{Towards the theory
  of reheating after inflation}, Phys. Rev. D 56 (1997) 3258--3295.
\newblock \href {https://doi.org/10.1103/PhysRevD.56.3258}
  {\path{doi:10.1103/PhysRevD.56.3258}}.
\newline\urlprefix\url{https://link.aps.org/doi/10.1103/PhysRevD.56.3258}

\bibitem{Greene-kofman1997}
P.~B. Greene, L.~Kofman, A.~Linde, A.~A. Starobinsky,
  \href{https://link.aps.org/doi/10.1103/PhysRevD.56.6175}{Structure of
  resonance in preheating after inflation}, Phys. Rev. D 56 (1997) 6175--6192.
\newblock \href {https://doi.org/10.1103/PhysRevD.56.6175}
  {\path{doi:10.1103/PhysRevD.56.6175}}.
\newline\urlprefix\url{https://link.aps.org/doi/10.1103/PhysRevD.56.6175}

\bibitem{kofman1996nonthermal}
L.~Kofman, A.~Linde, A.~A. Starobinsky,
  \href{https://link.aps.org/doi/10.1103/PhysRevLett.76.1011}{Nonthermal phase
  transitions after inflation}, Phys. Rev. Lett. 76 (1996) 1011--1014.
\newblock \href {https://doi.org/10.1103/PhysRevLett.76.1011}
  {\path{doi:10.1103/PhysRevLett.76.1011}}.
\newline\urlprefix\url{https://link.aps.org/doi/10.1103/PhysRevLett.76.1011}

\bibitem{Gottlober1991}
S.~Gottl\"ober, V.~M\"uller, A.~A. Starobinsky,
  \href{https://link.aps.org/doi/10.1103/PhysRevD.43.2510}{Analysis of
  inflation driven by a scalar field and a curvature-squared term}, Phys. Rev.
  D 43 (1991) 2510--2520.
\newblock \href {https://doi.org/10.1103/PhysRevD.43.2510}
  {\path{doi:10.1103/PhysRevD.43.2510}}.
\newline\urlprefix\url{https://link.aps.org/doi/10.1103/PhysRevD.43.2510}

\bibitem{Bruck2015}
C.~van~de Bruck, L.~E. Paduraru,
  \href{https://link.aps.org/doi/10.1103/PhysRevD.92.083513}{Simplest extension
  of starobinsky inflation}, Phys. Rev. D 92 (2015) 083513.
\newblock \href {https://doi.org/10.1103/PhysRevD.92.083513}
  {\path{doi:10.1103/PhysRevD.92.083513}}.
\newline\urlprefix\url{https://link.aps.org/doi/10.1103/PhysRevD.92.083513}


\bibitem{Planck2015}
P.~A.~R. Ade, N.~Aghanim, M.~Arnaud, M.~Ashdown, J.~Aumont, C.~Baccigalupi, and
  et~al.
\newblock {Planck 2015 results. XIII. Cosmological parameters}.
\newblock {\em Astron. Astrophys.}, 594:A13, 2016.
\newblock \href {http://arxiv.org/abs/1502.01589} {\path{arXiv:1502.01589}},
  \href {https://doi.org/10.1051/0004-6361/201525830}
  {\path{doi:10.1051/0004-6361/201525830}}.



\bibitem{Bruck2017}
C.~van~de Bruck, P.~Dunsby, L.~E. Paduraru,
  \href{https://doi.org/10.1142/S0218271817501528}{Reheating and preheating in
  the simplest extension of starobinsky inflation}, International Journal of
  Modern Physics D 26~(13) (2017) 1750152.
\newblock \href
  {http://arxiv.org/abs/https://doi.org/10.1142/S0218271817501528}
  {\path{arXiv:https://doi.org/10.1142/S0218271817501528}}, \href
  {https://doi.org/10.1142/S0218271817501528}
  {\path{doi:10.1142/S0218271817501528}}.
\newline\urlprefix\url{https://doi.org/10.1142/S0218271817501528}



\bibitem{Garcia_2023}
M.~A. Garcia, G.~Germán, R.~G. Quaglia, A.~M. Colorado,
  \href{https://dx.doi.org/10.1088/1475-7516/2023/12/015}{Reheating constraints
  and consistency relations of the starobinsky model and some of its
  generalizations}, Journal of Cosmology and Astroparticle Physics 2023~(12)
  (2023) 015.
\newblock \href {https://doi.org/10.1088/1475-7516/2023/12/015}
  {\path{doi:10.1088/1475-7516/2023/12/015}}.
\newline\urlprefix\url{https://dx.doi.org/10.1088/1475-7516/2023/12/015}




  \bibitem{Berera1996}
A.~Berera.
\newblock {Interpolating the Stage of Inflationary Universe: From Slow-roll to
  Warm Inflation}.
\newblock {\em Phys. Rev. D}, 54:2519--2534, 1996.
\newblock \href {http://arxiv.org/abs/hep-th/9601134}
  {\path{arXiv:hep-th/9601134}}, \href
  {https://doi.org/10.1103/PhysRevD.54.2519}
  {\path{doi:10.1103/PhysRevD.54.2519}}.


  \bibitem{Berera2005}
  Arjun Berera and Rudnei~O. Ramos.
  \newblock Dynamics of interacting scalar fields in expanding space-time.
  \newblock {\em Phys. Rev. D}, 71:023513, Jan 2005.
  \newblock URL: \url{https://link.aps.org/doi/10.1103/PhysRevD.71.023513}, \href
  {https://doi.org/10.1103/PhysRevD.71.023513}
  {\path{doi:10.1103/PhysRevD.71.023513}}.

\bibitem{Saha2026}
Bhargabi Saha and Malay K. Nandy.
\newblock {Dynamical evolution and graceful exit in quartic warm inflation}.
\newblock {\em Journal of Cosmology and Astroparticle Physics}, 2026:044, 2026.
\newblock \href {https://doi.org/10.1088/1475-7516/2026/05/044}
  {\path{doi:10.1088/1475-7516/2026/05/044}}.



\bibitem{Saha2025}
Bhargabi Saha and Malay K. Nandy.
\newblock {The $\alpha$-attractor $E$-model in warm inflation: observational viability from Planck 2018}.
\newblock {\em Journal of Cosmology and Astroparticle Physics}, 2025:102, 2025.
\newblock \href {https://doi.org/10.1088/1475-7516/2025/10/102}
  {\path{doi:10.1088/1475-7516/2025/10/102}}.



\bibitem{Kallosh2013a}
Renata Kallosh, Andrei Linde, and Diederik Roest.
\newblock Superconformal inflationary $\alpha$-attractors.
\newblock {\em JHEP}, 11:198, 2013.
\newblock \href {http://arxiv.org/abs/1311.0472} {\path{arXiv:1311.0472}},
  \href {https://doi.org/10.1007/JHEP11(2013)198}
  {\path{doi:10.1007/JHEP11(2013)198}}.

 \bibitem{Kallosh2013b}
Renata Kallosh and Andrei Linde.
\newblock Superconformal generalizations of the Starobinsky model.
\newblock {\em Journal of Cosmology and Astroparticle Physics}, 2013(06):028,
  2013.
\newblock URL:
  \url{https://iopscience.iop.org/article/10.1088/1475-7516/2013/06/028}, \href
  {https://doi.org/10.1088/1475-7516/2013/06/028}
  {\path{doi:10.1088/1475-7516/2013/06/028}}.



\bibitem{Linde2014}
Renata Kallosh, Andrei Linde,  and Alexander Westphal,
\newblock {Chaotic inflation in supergravity after Planck and BICEP2}.
\newblock {\em Phys. Rev. D}, 90:023534, 2014.
  \href {https://link.aps.org/doi/10.1103/PhysRevD.90.023534}
  {\path{10.1103/PhysRevD.90.023534}}.



  \bibitem{Linde2025}
Renata Kallosh and Andrei Linde.
\newblock {On the Present Status of Inflationary Cosmology}
(Proceedings of the Lema\^itre Conference 2024
``Black Holes, Gravitational Waves and Space-Time Singularities'').
\newblock {\em arXiv preprint: hep-th 2505.13646},  (2025).
\newblock \href {https://arxiv.org/abs/2505.13646}
  {\path{arxiv.org/abs/2505.13646}}.


  \bibitem{Planck2018}
Y. Akrami, et~al.
\newblock Planck 2018 results. X. Constraints on inflation.
\newblock {\em Astron. Astrophys.}, 641:A10, 2020.
\newblock \href {http://arxiv.org/abs/1807.06211} {\path{arXiv:1807.06211}},
  \href {https://doi.org/10.1051/0004-6361/201833887}
  {\path{10.1051/0004-6361/201833887}}.


  \bibitem{kofman_PRD_2005}
  L.~Kofman, P.~Yi,
  \href{https://link.aps.org/doi/10.1103/PhysRevD.72.106001}{Reheating the
  universe after string theory inflation}, Phys. Rev. D 72 (2005) 106001.
\newblock \href {https://doi.org/10.1103/PhysRevD.72.106001}
  {\path{doi:10.1103/PhysRevD.72.106001}}.
\newline\urlprefix\url{https://link.aps.org/doi/10.1103/PhysRevD.72.106001}


\bibitem{albrecht1982reheating}
A.~Albrecht, P.~J. Steinhardt, M.~S. Turner, F.~Wilczek,
  \href{https://link.aps.org/doi/10.1103/PhysRevLett.48.1437}{Reheating an
  inflationary universe}, Phys. Rev. Lett. 48 (1982) 1437--1440.
\newblock \href {https://doi.org/10.1103/PhysRevLett.48.1437}
  {\path{doi:10.1103/PhysRevLett.48.1437}}.
\newline\urlprefix\url{https://link.aps.org/doi/10.1103/PhysRevLett.48.1437}


\bibitem{Whitt1984}
B.~Whitt,
  \href{https://www.sciencedirect.com/science/article/pii/0370269384903320}{Fourth-order
  gravity as general relativity plus matter}, Physics Letters B 145~(3) (1984)
  176--178.
\newblock \href {https://doi.org/https://doi.org/10.1016/0370-2693(84)90332-0}
  {\path{doi:https://doi.org/10.1016/0370-2693(84)90332-0}}.
\newline\urlprefix\url{https://www.sciencedirect.com/science/article/pii/0370269384903320}



\bibitem{Faulkner_2007}
T.~Faulkner, M.~Tegmark, E.~F. Bunn, Y.~Mao,
  \href{https://link.aps.org/doi/10.1103/PhysRevD.76.063505}{Constraining
  $f(r)$ gravity as a scalar-tensor theory}, Phys. Rev. D 76 (2007) 063505.
\newblock \href {https://doi.org/10.1103/PhysRevD.76.063505}
  {\path{doi:10.1103/PhysRevD.76.063505}}.
\newline\urlprefix\url{https://link.aps.org/doi/10.1103/PhysRevD.76.063505}


\bibitem{Komatsu_2009}
E.~Komatsu, J.~Dunkley, M.~R. Nolta, C.~L. Bennett, B.~Gold, G.~Hinshaw,
  N.~Jarosik, D.~Larson, M.~Limon, L.~Page, D.~N. Spergel, M.~Halpern, R.~S.
  Hill, A.~Kogut, S.~S. Meyer, G.~S. Tucker, J.~L. Weiland, E.~Wollack, E.~L.
  Wright, \href{https://dx.doi.org/10.1088/0067-0049/180/2/330}{Five-year
  wilkinson microwave anisotropy probe* observations: Cosmological
  interpretation}, The Astrophysical Journal Supplement Series 180~(2) (2009)
  330.
\newblock \href {https://doi.org/10.1088/0067-0049/180/2/330}
  {\path{doi:10.1088/0067-0049/180/2/330}}.
\newline\urlprefix\url{https://dx.doi.org/10.1088/0067-0049/180/2/330}


 \bibitem{Starobinsky1984}
A.~A.~Starobinsky, ``Nonsingular Model of the Universe with the
Quantum-Gravitational de Sitter Stage and its Observational Consequences,''
in {\em Quantum Gravity}, edited by M.~A. Markov and P.~C. West (Springer,
Boston, MA, 1984) pp. 103--128.
\newblock \href {https://doi.org/10.1007/978-1-4613-2701-1_8}
  {\path{doi:10.1007/978-1-4613-2701-1_8}}.


\bibitem{d2017universe}
F.~D'Eramo, N.~Fernandez, S.~Profumo,
  \href{https://dx.doi.org/10.1088/1475-7516/2017/05/012}{When the universe
  expands too fast: relentless dark matter}, Journal of Cosmology and
  Astroparticle Physics 2017~(05) (2017) 012.
\newblock \href {https://doi.org/10.1088/1475-7516/2017/05/012}
  {\path{doi:10.1088/1475-7516/2017/05/012}}.
\newline\urlprefix\url{https://dx.doi.org/10.1088/1475-7516/2017/05/012}


\end{thebibliography}

\end{document}